\documentclass{ws-ijmpd}
\usepackage[super,compress]{cite}
\usepackage{color}
\usepackage{url}
\begin{document}

\markboth{R. Ciolfi}
{Short GRB central engines}

%
\catchline{}{}{}{}{}
%

\title{SHORT GAMMA-RAY BURST CENTRAL ENGINES}

\author{RICCARDO CIOLFI}

\address{INAF, Osservatorio Astronomico di Padova, Vicolo 
dell'Osservatorio 5, I-35122 Padova, Italy\\
\vspace{0.1cm}
INFN-TIFPA, Trento Institute for Fundamental Physics
and Applications,\\ Via Sommarive 14, I-38123 Trento, Italy\\
\vspace{0.1cm}
riccardo.ciolfi@inaf.it}

\maketitle

\begin{history}
\received{Day Month Year}
\revised{Day Month Year}
\end{history}

\begin{abstract}
Growing evidence connects the progenitor systems of the short-hard subclass of gamma-ray bursts (GRBs) to the merger of compact object binaries composed by two neutron stars (NSs) or by a NS and a black hole (BH). 
The recent observation of the binary NS (BNS) merger event GW170817 associated with GRB 170817A brought a great deal of additional information and provided further support to the above connection, even though the identification of this burst as a canonical short GRB (SGRB) remains uncertain.
Decades of observational constraints and theoretical models consolidated the idea of a jet origin for the GRB prompt emission, which can also explain the multiwavelength afterglow radiation observed in most of the events. 
However, the mechanisms through which a BNS or NS-BH merger remnant would power a collimated outflow are much less constrained.
Understanding the properties of the remnant systems and whether they can provide the right conditions for jet production has been a main driver of the great effort devoted to study BNS and NS-BH mergers, and still represents a real challenge from both the physical and the computational point of view.
One fundamental open question concerns the nature of the central engine itself. 
While the leading candidate system is a BH surrounded by a massive accretion disk, the recent observation of plateau-shaped X-ray afterglows in some SGRBs would suggest a longer-lived engine, i.e.~a metastable (or even stable) massive NS, which would also exclude NS-BH progenitors.
Here we elaborate on this key aspect, considering three different scenarios to explain the SGRB phenomenology based on different hypotheses on the nature of the merger remnant. 
Then, we discuss the basic properties of GRB 170817A and how this event would fit within the different frameworks of the above scenarios, under the assumption that it was or was not a canonical SGRB. 
\end{abstract}


\ccode{PACS numbers: 04.25.D-, 04.25.dk, 04.30.Tv, 04.40.Dg, 98.70.Rz}


\section{Introduction}
\label{sec:intro}

Gamma-ray bursts (GRBs) are bright non-repeating flashes of radiation observed in the gamma-ray band at a current rate of the order of a few hundreds per year, characterized by randomly distributed sky positions, a duration from a fraction of a second up to hundreds or even thousands of seconds, and a diversity of photometric and spectral properties. 
Their extragalactic origin is well established, placing these events at cosmological distances and implying isotropic-equivalent luminosities up to $\sim$\,10$^{54}$\,erg/s (among the most luminous phenomena known in the Universe).  
Moreover, they are often followed by afterglow signals across the electromagnetic spectrum from X-ray to radio wavelengths and covering timescales from tens of seconds up to several months or more. 

Besides the wide range of temporal and spectral properties, a bimodal distribution based on duration and spectral hardness has been identified: GRBs are now commonly divided into short-hard and long-soft bursts,\cite{Kouveliotou1993} with a separation around $T_{90}\approx2$~s (where $T_{90}$ is time window containing 90\% of the energy released). 
This bimodal distribution is indicative of two distinct types of GRBs, associated with different progenitor systems. Two key observational results allowed to relate long GRBs (conventionally, those having $T_{90}>2$~s) with the death of massive stars (``collapsar" model\cite{MacFadyen1999}). First, these events have been observed only in star-forming galaxies and often found in the brightest star-forming regions within the host galaxy itself (e.g., Ref.~\refcite{Bloom1998,Fruchter2006}). Second and most importantly, a number of direct associations of long GRBs with type Ic core-collapse supernovae have been collected over the last 20 years (starting with SN 1998bw\cite{Galama1998}), removing previous doubts on the connection between the two phenomena.
Conversely, the progenitor system responsible for short GRBs (SGRBs) has not been confidently identified and an association with core-collapse supernovae seems to be excluded by the observations.

According to the leading explanation, SGRBs are powered by the merger of compact binary systems composed of two neutron stars (NSs) or a NS and a black hole (BH). This idea was already put forward before the GRB bimodal distribution was firmly assessed, based on broadly consistent event rates and the fact that these systems were compatible with short-duration and powerful central engines, $\sim$\,ms dynamical timescales, and baryon-poor environments\cite{Paczynski1986,Eichler:1989:126,Narayan1992}. 
Mounting (indirect) evidence further strenghtened this idea and, at the same time, disfavoured the connection with core-collapse supernovae. This includes the lack of supernova associations, the observation in all type of host galaxies (also elliptical galaxies with very low star formation), and the larger offsets from the center of the host galaxies, indicative of progenitor systems with large natal kicks and a significant time delay between their formation and the SGRB event (see, e.g., Ref.~\refcite{Berger2014} and Refs.~therein).

An additional indication in support to the above scenario was provided by the observation of an optical/IR rebrightening in the afterglow emission of the short GRB 130603B, which was tentatively interpreted as a signature of a binary merger\cite{Tanvir2013,Berger2013}. 
Binary NS (BNS) and NS-BH mergers are expected to eject a significant amount of hot and neutron-rich material, either dynamically along with the merger process or after merger, via baryon-loaded winds from the remnant system (possibly a BH surrounded by an accretion disk).  
Since the early predictions that r-process nucleosynthesis of heavy elements would take place in these conditions\cite{Lattimer1974,Lattimer1977,Eichler:1989:126} and that the radioactive decay of these elements could lead to potentially observable thermal emission\cite{Li:1998:L59}, it has been clear that the resulting signal, named ``kilonova" (or ``macronova"), should be regarded among the possible electromagnetic counterparts to be expected from compact binary mergers along with their gravitational wave (GW) emission (e.g., Ref.~\refcite{Metzger2012}). The optical/IR rebrightening of GRB 130603B was found consistent with a kilonova, suggesting that this burst coincided with a merger event.

Combined together, the above indications make a very strong case for the merger progenitor hypothesis, though not conclusive.
A smoking gun evidence would be represented by a SGRB detection in coincidence with the GW signal from the merger. 
The first detection of GWs from a BNS merger\footnote{We assume here it was a BNS, although a NS-BH binary cannot be completely excluded based on the GW observation only.} on August 17$^\mathrm{th}$ 2017 (event named GW170817) by the LIGO Scientific Collaboration and the Virgo Collaboration{\cite{LVC-BNS} was accompanied by the observation of multiwavelength electromagnetic emission across the entire spectrum, from gamma-rays to radio, marking also the first multimessenger observation of a GW source{\cite{LVC-MMA}. While the optical/IR signal provided the first compelling evidence for a radioactively-powered kilonova and confirmed the key role played by BNS mergers in the production of heavy elements in the Universe (e.g., Ref.~\refcite{Cowperthwaite2017,Kasen2017,Kasliwal2017,Pian2017,Tanvir2017,Smartt2017}, among others), a short-duration gamma-ray signal (GRB 170817A) emerging less than 2 s after merger\cite{LVC-GRB,Goldstein2017} combined with late-time X-ray, radio, and optical afterglows indicated the presence of a mildly relativistic outflow interacting with the interstellar medium (e.g., Ref.~\refcite{Troja2017,Margutti2017,Haggard2017,Hallinan2017,Mooley2018,Lazzati2018,Margutti2018,Nakar2018}, among others). 
If this event produced a SGRB, it would represent the smoking gun evidence we have been waiting for. Nevertheless, as we discuss in a devoted section of this Paper (Section \ref{GRB170817}), the properties of the observed gamma-ray signal combined with other elements (e.g., the viewing angle) still leave some doubts on whether the merger was accompanied by a ``canonical'' SGRB (i.e.~a GRB that would have looked like the other known SGRBs if seen with the right viewing angle). 

For both types of GRBs, the non-thermal spectra and short variability timescales of the prompt gamma-ray emission, combined with the huge energy release, point to a highly relativistic outflow with bulk Lorentz factors $\Gamma\gtrsim10^2$ and very low baryon loading{\cite{Paczynski1986}. Decades of investigation converged to the idea that GRBs are associated with relativistic jets, where most of the outflow is initially collimated within a relatively small opening angle (from a few to about 10 degrees). The jet framework also explains in a natural way the afterglow emission often observed at lower energies, from X-rays to radio, as the result of the jet interaction with the interstellar medium and the consequent production of synchrotron radiation at the forward-shock (e.g., Ref.~\refcite{Rees1992}). This view is further supported by the detection of an achromatic steepening of the decaying lightcurve in many GRB afterglows (``jet break''), which offered a way to directly infer the jet opening angles\cite{Rhoads1999,Sari1999}.
Despite the wide consensus on the above broad picture, the actual GRB phenomenology is much more complex, with many open questions on the interpretation of both the prompt and the afterglow radiation (see, e.g., Ref.~\refcite{Kumar2015} for a recent review on GRBs and jets). 
The mechanism responsible for the non-thermal gamma-ray emission, for instance, is not fully understood. The leading model associates the prompt emission with internal shocks produced within the jet by the collision of outflow shells launched by the central engine with different Lorentz factors, which results in strong particle acceleration{\cite{Rees1994}. A possible alternative is that the gamma-ray emission has instead a photospheric origin (e.g., Ref.~\refcite{Rees2005,Giannios2006,Lazzati2009}). The jet composition itself is also unclear, with the main energy reservoir possibly carried by baryonic matter, electron-positron pairs, or magnetic fields (e.g., Ref.~\refcite{Kumar2015}).

The interpretation of GRB data is based on the assumption that the observed GRBs are seen on-axis, i.e.~along a direction contained within the opening angle of the jet. This leaves an open question on how the GRB outflow is structured at angles larger than $\theta_\mathrm{jet}$ (i.e.~outside the high Lorentz factor region) and which kind of signal would be detected by an off-axis observer ($\theta_\mathrm{obs}>\theta_\mathrm{jet}$). As discussed in Section~\ref{GRB170817}, the answer to this question is key to understand the nature of GRB 170817A, the peculiar burst accompanying the BNS merger event GW170817. 

If we assume a merger origin for SGRBs and based on our current understanding of their prompt and afterglow emission, we have to conclude that BNS and/or NS-BH mergers
can launch relativistic jets with properties matching the observations, in terms of energy, terminal Lorentz factor, duration, variability, collimation, and so on.
However, no direct proof exists on theoretical grounds and strong uncertainties remain on both the underlying physical mechanisms and the nature of the central engine itself, which could be either a BH surrounded by an accretion disk or a long-lived massive NS. This represents the most challenging and least understood aspect of the connection between SGRBs and BNS/NS-BH mergers.

In this Paper, we address the above fundamental issue by elaborating on different SGRB central engine scenarios proposed in the literature. 
In particular, we focus the attention on three main scenarios: (i) the most popular BH-accretion disk scenario, (ii) the magnetar scenario, (iii) and the so-called ``time-reversal'' scenario. 
Note that most of the current investigation on SGRBs refers to the first two, while the third was proposed only recently and is further developed here (Section~\ref{TR}).  
Our discussion takes into account both the current observational constraints and the most recent theoretical advancements (including the latest results from numerical simulations of BNS/NS-BH mergers), 
with the aim of drawing a comprehensible and up-to-date picture where the strengths and weaknesses of each scenario are clearly identified. This is the subject of Section~\ref{engine}.
Then, in Section~\ref{GRB170817}, we turn to consider the specific properties of GRB 170817A. This peculiar event may belong to the known class of SGRBs or possibly represent a new/distinct type of GRB. In either cases, we discuss its compatibility with our three reference scenarios. 
Finally, in Section~\ref{conclusion}, we present a summary of the Paper and our concluding remarks. 


\section{Central engine scenarios for short gamma-ray bursts}
\label{engine}

A growing theoretical effort is devoted to study the possible outcomes of BNS and NS-BH mergers and to
establish the capability of the remnant system to launch a jet and thus act as the central engine powering a SGRB. 
Numerous GRB observations provided direct indications on the jet structure and evolution, as well as important constraints on the possible mechanisms responsible for the prompt gamma-ray emission and the multiwavelength afterglows. However, much less observational constraints exist on the nature of the central engine, on the launching mechanism, or on the initial properties of the incipient jet close to the launching region. This lack of information, combined with the very rich and partially unknown physics involved in the merger process and in the post-merger evolution, makes this investigation extremely challenging. 

One more reason for the present difficulties comes from the computational limitations. 
General relativistic (magneto)hydrodynamics simulations including all the key physical ingredients (e.g., tabulated nuclear physics equations of state, neutrino emission and transport, magnetic fields) are necessary in order to properly describe the merger and obtain solid conclusions
(see, e.g., Ref.~\refcite{Paschalidis2017,Baiotti2017} for recent reviews). 
Currently, adaptive mesh refinement with a minimum grid spacing of $\Delta x\approx100-200$\,m (covering a NS radius with about 50\,$-$\,100 points) is commonly employed. On the one hand, this is still insufficient to fully resolve important small-scale effects associated with turbulence, effective viscosity, and magnetic field amplification mechanisms (e.g., the Kelvin-Helmholtz instability\cite{Kiuchi:2015:1509.09205}). On the other hand, the Courant limitation imposes a corresponding timestep as small as $\sim$\,10$^{-4}$\,ms. 
As a result, typical simulations, despite being computationally very expensive, cover no more than $\sim$\,100\,ms of evolution, which is only sufficient to follow the last few orbits of inspiral until merger and further evolve the post-merger system for several tens of ms. 
Moreover, the overall computational domain extends up to a few thousand kilometers at most. 
The goal of these simulations is to show that an incipient jet can be launched for specific conditions and via a specific mechanism, but they cannot provide the whole picture. 
The evolution of an incipient jet on time and spatial scales larger by orders of magnitude is however crucial to define its ultimate properties (luminosity, collimation, terminal Lorentz factor, and so on) and thus to connect the simulation results to observable features. 
These larger scales are currently covered by sophisticated special relativistic simulations, which start from hand-made initial conditions possibly inspired by the results of merger simulations (e.g., Ref.~\refcite{Nagakura2014,Murguia-Berthier2014,Murguia-Berthier2017a,Lazzati2017,Gottlieb2018}). At present, there is no proper match between the two types of simulations that could allow for a self-consistent description of the entire evolution. 

In this Section (\ref{BH}\,$-$\,\ref{TR}), we discuss separately the status of the theoretical investigation within three different scenarios, which differ for the nature of the merger remnant.
For NS-BH mergers, the remnant has to be a BH, possibly surrounded by an accretion disk. 
As we further discuss in the next Section (\ref{BH}), a BH-disk system has in principle all the necessary ingredients to launch an accretion-powered relativistic jet, provided that the disk is massive enough and the accretion lasts for a sufficiently long time. 
For a given NS mass and a given hypothesis on the NS equation of state (EOS), which determines its compactness, the above requirements are met only for a limited range of BH masses and spins (e.g., Ref.~\refcite{Foucart2012,Pannarale2014}), such that the NS is tidally disrupted outside the BH innermost stable circular orbit (ISCO). In particular, larger BH masses and smaller BH spins make this condition more difficult to satisfy, resulting in a limited fraction of NS-BH systems that would be able to power a jet.

The phenomenology of BNS mergers is more rich. Most of these mergers will eventually end up in BH, but only after an intermediate state consisting of a massive metastable NS remnant. The survival time of such a remnant prior to collapse depends mostly on total mass, mass ratio between the two NSs, and the NS EOS, but it can also be sensitive to physical effects becoming dynamically important after merger, like magnetic fields and neutrino cooling. 
A common set of definitions based solely on the remnant mass and referring to a given EOS separates the possible outcomes into hypermassive NSs (HMNSs), with mass above the maximum mass allowed for a uniformly rotating configuration, supramassive NSs (SMNSs), with mass below this limit but still above the maximum mass allowed for a nonrotating configuration, and ``stable'' NSs, with mass below the latter limit. 

A HMNS can only survive in presence of strong differential rotation and will typically collapse on timescales $\lesssim$\,100\,ms. 
In this case, an upper limit to the lifetime is dictated by the timescale for removal of differential rotation in the NS core (via magnetic fields and/or other mechanisms). Nevertheless, higher mass will lead to an earlier collapse, down to less than a few ms (in which case we talk about prompt collapse). The final system is a spinning BH surrounded by a disk of variable mass. Very high NS mass may lead to a disk of negligible mass, while more typical disk masses reach $\sim$\,0.1\,$M_\odot$.

The lifetime of a SMNS is potentially much longer. By definition, uniform rotation can support the NS against collapse as long as the centrifugal support is sufficient for the given mass. This may result in survival times comparable to the timescale associated with the dominant spindown mechanism, which is typically GW emission in the first tens of ms and at later times, if the remnant has not collapsed yet,  magnetic dipole radiation. For typical dipolar magnetic field strengths expected at the pole of the remnant NS ($\sim$\,10$^{14}-10^{15}$\,G) the corresponding lifetime can be as long as several hours. Note, however, that a high mass SMNS can also collapse on much shorter timescales, down to $\sim$\,1\,s or even less. 
The lifetime of a SMNS is essentially unknown for cases where the collapse happens before uniform rotation is achieved, due to the complex structure and dynamics of the remnant in such early phases. This also leaves doubts on the possibility that in some cases a SMNS might end up in a BH surrounded by a massive disk. We further elaborate on this point in Section~\ref{TR}. 

Finally, a stable NS will never collapse to a BH. It will first achieve uniform rotation and then continue its spindown evolution forever. In what follows, we will use the term {\it long-lived NS} referring to any remnant with a lifetime $\gtrsim$\,1\,s, thus including both SMNSs and stable NSs. 

The recent observation of single NSs with a mass of $\approx$\,2\,$M_\odot$\cite{Demorest2010,Antoniadis2013} excluded the softest NS EOS proposed so far and contributed to question the previous notion that a BNS merger will almost always produce a HMNS. The predicted distribution of NS masses in a merging BNS system is sharply peaked around 1.3\,$-$\,1.4\,$M_\odot$ (e.g., Ref.~\refcite{Belczynski2008}) and thus the typical remnant (gravitational) mass, once binding energy, neutrino losses, and mass ejection are taken into account, should be $\sim$\,2.5\,$M_\odot$. Knowing that a slowly rotating NS can support more than about 2\,$M_\odot$ and that masses up to $\sim$\,20\% higher (i.e. $\gtrsim$\,2.4\,$M_\odot$) can be supported in presence of the rapid ($\sim$\,kHz) rotation expected after merger (e.g., Ref.~\refcite{Lasota1996}), we conclude that a long-lived NS should be a rather likely outcome of BNS mergers (e.g., Ref.~\refcite{Piro2017,Gao2016}).

For the first BNS merger observed in August 2017, the estimated masses of the two NSs fit very well within the range of masses and mass ratios that were expected for a typical merger\cite{LVC-BNS}. 
Unfortunately, this event did not provide a direct and clear indication on the nature of the remnant, but for most EOS the remnant could have been either a HMNS or a SMNS (e.g., Ref.~\refcite{LVC-GRB}) . 
In Section~\ref{GRB170817}, we consider some indications on the nature of the remnant coming from the comparison with the different SGRB scenarios and their viability, also depending on the assumed nature of GRB 170817A. In this Paper, however, we do not provide a full discussion on the constraints or indications obtained from the analysis of the GW or the kilonova signals. 


\subsection{Black hole-accretion disk scenario}
\label{BH}

A BH surrounded by a massive accretion disk is a very natural and perhaps the most likely outcome of a BNS merger. BHs are also the necessary outcome of NS-BH mergers and there is a realistic range of BH masses and spins for which a significant amount of the NS matter ($\gtrsim$\,0.1\,$M_\odot$) is retained in the accretion disk formed after merger.
BH-disk systems like these, characterized by huge accretion rates (up to $\dot{M}$\,$\sim$\,0.1\,$-$\,1\,$M_\odot/$s, orders of magnitude above the Eddington limit), are well known to be the potential source of a relativistic jet, which in turn makes them ideal candidates for the central engine powering SGRBs. 

According to the most popular scenario (hereafter ``BH-disk'' scenario) a SGRB jet is indeed launched by a hyperaccreting stellar-mass BH produced in a BNS or a NS-BH merger\cite{Eichler:1989:126,Narayan1992,Mochkovitch1993}. 
Besides the agreement with the indirect evidence suggesting a merger origin for SGRBs, these systems offer an almost baryon-free environment along the BH spin axis that is suitable for launching a relativistic outflow and can in principle reproduce all the right scales. 
The prompt SGRB emission, once corrected for a typical jet opening angle of 
$\theta_\mathrm{jet}$\,$\sim$\,10$^{\circ}$, corresponds to a total energy of 
$E_\mathrm{prompt}$\,$\sim$\,10$^{49}-10^{50}$\,erg released in less than 2 seconds, resulting in luminosities of the order of $L_\mathrm{prompt}$\,$\sim$\,10$^{49}-10^{50}$\,erg/s. 
Accretion timescales around BHs of mass in the range $3-10$\,$M_\odot$ (NS-BH mergers leading to $M_\mathrm{BH}$\,$\gtrsim$10\,$M_\odot$ are excluded as they will hardly produce a massive accretion disk) are of the order of $\tau_\mathrm{accr}$\,$\sim$\,0.1$-1$\,s, as confirmed by numerical relativity simulations. This is consistent with the requested duration of the SGRB central engine activity.
Taking a reference disk mass of $M_\mathrm{disk}$\,$\sim$\,0.1\,$M_\odot$ we obtain $E_\mathrm{prompt}/M_\mathrm{disk}c^2$\,$\sim$\,10$^{-4}-10^{-3}$ and 
$\eta$\,$\equiv$\,$L_\mathrm{prompt}/\dot{M}c^2$\,$\sim$\,10$^{-4}-10^{-2}$.
Therefore, a mechanism able to convert the accretion rate into SGRB luminosity with an efficiency of up to $\eta$\,$\sim$\,1\% would potentially explain the observations. 
Note that for smaller disk masses the requirements on $\eta$ become more demanding. This is why a disk mass $\gtrsim$\,0.1\,$M_\odot$, which is commonly found in merger simulations, is usually considered necessary to power SGRB jets.  

Two leading mechanisms have been proposed as the possible source of energy powering a jet: neutrino-antineutrino ($\nu\bar{\nu}$) annihilation along the BH spin axis\cite{Eichler:1989:126} and magnetohydrodynamic effects possibly involving the Blandford-Znajek mechanism\cite{Blandford1977}. Note that both mechanisms can be at play simultaneously, although so far they have only been considered separately under the assumption that one of the two is dominant. 

Massive disks around hyperaccreting stellar-mass BHs are thick (i.e. of toroidal shape), rather compact, and reach very high temperatures (up to $\sim$\,MeV). In these conditions, copious neutrino and antineutrino emission is expected especially from the inner part of the disk, attaining initial luminosities of $L_\nu$\,$\sim$\,10$^{52}$\,erg/s rapidly decreasing over the accretion timescale (e.g., Ref.~\refcite{Ruffert:1999:573,Just2016}). While this emission can amount to a few percent of $\dot{M}c^2$, the rate of energy deposition at the poles of the BH via $\nu\bar{\nu}$ annihilation is at least one order of magnitude lower. Recent simulations find $\sim$\,10$^{49}$\,erg of cumulative annihilation energy (e.g., Ref.~\refcite{Just2016,Perego2017}), which is at the lower end of the minimal energy budget necessary to explain SGRB jets. Even assuming that most of this energy will be converted into escaping radiation, these results show significant tension with the hypothesis that a pure neutrino mechanism would be sufficient to power most (if not all) SGRB jets.

An additional potential obstacle for the production of a successful jet is the baryon-pollution along the spin axis of the BH. 
In the BNS case, the surroundings of the merger site can be polluted by both (i) dynamical ejecta launched during the merger itself and (ii) post-merger baryon-loaded winds expelled by the accretion disk and possibly by the metastable NS remnant before its collapse. 
Dynamical ejecta have a tidally-driven contribution due to tidal effects acting on the NS(s) during the very last orbit of the inspiral, which is mostly confined to the equatorial plane, and a shock-driven contribution associated with shocks produced by the two NS cores crashing into each other and by the first most violent oscillations of the NS remnant, which is instead mostly polar (e.g., Ref.~\refcite{Hotokezaka:2013:24001,Ciolfi2017}).
Post-merger winds are nearly isotropic and driven by the effective pressure associated with neutrino emission and reabsorption and/or magnetic fields (e.g., Ref.~\refcite{Dessart:2009:1681,Siegel:2014:6,Siegel2017}).
When the central engine activity starts and the incipient jet is formed, the latter has to drill through the post-merger winds already expelled plus the shock-driven dynamical ejecta. 
Simulations studying this jet-ejecta interaction seem to converge to the conclusion that a neutrino-powered jet would have difficulties in successfully emerging from such a polluted environment, and even when this would happen the jet energy would be likely insufficient to explain a typical SGRB (e.g., Ref.~\refcite{Murguia-Berthier2014,Just2016}).

For NS-BH mergers, dynamical ejecta are only tidally-driven and equatorial and post-merger matter outflows are only those expelled by the accretion disk. Therefore, the amount of ejecta along the polar direction is minimal when the incipient jet is supposed to be launched. This makes baryon pollution less restrictive for the NS-BH case. Nevertheless, a jet powered only by $\nu\bar{\nu}$ annihilation would only be sufficient to explain SGRBs of relatively low luminosity.

The above difficulties of the neutrino mechanism suggest that magnetic fields should play a major role in powering SGRB jets.
Although there exist different magnetohydrodynamic mechanisms in which a BH-disk system could launch a jet (e.g., Ref.~\refcite{Meier2003,Abramowicz2013}), the most discussed (and perhaps most promising) in the context of BNS and NS-BH mergers is the Blandford-Znajek (BZ) mechanism\cite{Blandford1977}.
By means of this mechanism, a spinning (Kerr) BH threaded by a strong magnetic field connected to an external load of material (i.e.~a magnetized accretion disk) can power a Poynting-flux dominated outflow at the expense of its own rotational energy, which is effectively extracted via magnetic torque.
The resulting Poynting-flux luminosity is approximately given by\cite{Thorne1986}
\begin{equation}
L_\mathrm{BZ}\sim10^{51} (\chi/0.8)^2 (M_\mathrm{BH}/6\,M_\odot)^2 (B_\mathrm{BH}/10^{15}\,\mathrm{G})^2~\mathrm{erg/s}\, ,
\label{BZ1}
\end{equation}
where $\chi$ and $M_\mathrm{BH}$ are the BH dimensionless spin and mass, while $B_\mathrm{BH}$ is the characteristic magnetic field strength close to the BH. The reference values in Eq.\,(\ref{BZ1}) are indicative for a NS-BH merger. Using $\chi$\,$=$\,0.6 and $M_\mathrm{BH}$\,$=$\,2.5\,$M_\odot$, more typical for BNS mergers, would give $L_\mathrm{BZ}$\,$\sim$\,10$^{50}$\,erg/s.
By directly tapping the rotational energy of the BH, the available energy reservoir can be very high. However, in order to be sufficient to explain the prompt emission of SGRBs, strong magnetic fields $B_\mathrm{BH}$\,$\gtrsim$\,10$^{15}$ are also necessary. 
The BZ process has been already seen at work in numerical simulations. In particular, in the context of AGNs efficiencies $\eta_\mathrm{BZ}$\,$\equiv$\,$L_\mathrm{BZ}/\dot{M}c^2$\,$>$\,100\,\% have been reported, demonstrating that power is actually extracted from the BH rotational energy\cite{Tchekhovskoy2012}. 

Hydrodynamics simulations of NS-BH mergers in full general relativity have been succesfully performed for over a decade (see, e.g., Ref.~\refcite{Shibata2011} for a review) and provided important indications on the viability of these systems as SGRB central engines. In particular, they allowed to define the range of masses and spins of the initial BH and the NS companion that would lead to a massive accretion disk ($M_\mathrm{disk}$\,$\gtrsim$\,0.1\,$M_\odot$). For quasi-circular binaries with a nonspinning (i.e.~irrotational) NS and $M_\mathrm{BH}$ in the realistic range 4\,$-$\,10\,$M_\odot$, high BH spins of $\chi$\,$\gtrsim$\,0.8 are necessary for most EOS\cite{Foucart2012}.
More compact NSs or more massive BHs make this requirement even more tight. In presence of eccentricity (e.g.~for binaries formed in globular clusters via capture or exchange interactions), the requirements on the BH spin are instead less stringent\cite{East2015}. 

Due to the complexity associated with the inclusion of magnetic fields, however, general relativistic magnetohydrodynamics (GRMHD) simulations of NS-BH mergers that could directly explore the potential to form jets via the BZ mechanism are not as many (e.g., Ref.~\refcite{Chawla2010,Etienne2012a,Etienne2012b,KiuchiSek2015,Paschalidis2015}).
These simulations were able to show the formation of an ordered poloidal magnetic field structure along the BH axis, but the formation of an incipient jet was only reported in Ref.~\refcite{Paschalidis2015}.
In this case, the authors found the emergence of a magnetically-dominated and mildly relativistic outflow inside the baryon-free funnel formed along the BH spin axis. This outflow emerged $\sim$\,100\,ms after merger and the Poynting-flux luminosity matched well the expected $L_\mathrm{BZ}$ given in Eq.\,(\ref{BZ1}). 
Such a result was only found for simulations starting with an initial NS magnetic field having a poloidal component that extended also outside the NS and thus threading the BH. Adopting an initial field entirely confined inside the NS, as in all previous simulations, gave no incipient jet.  
Another key issue is the initial magnetic field strength. The authors imposed a maximum magnetic field strength inside the NS of $\sim$\,10$^{17}$\,G, orders of magnitude higher than the typical $\sim$\,10$^{12}$\,G (at the pole) inferred from pulsar observations. As a consequence, this encouraging result can only be taken as a proof of principle on the viability of NS-BH merger engines for SGRBs.
Nevertheless, strong magnetic field amplification is expected to take place in the accretion disk, mainly via the magnetorotational instability (MRI),\cite{Balbus:1991,Etienne2012a} and thus much weaker initial magnetic fields could grow in principle up the levels required for SGRB central engines. 
Current resolutions are insufficient to properly resolve amplification mechanisms acting on small-scales (such as the MRI), but a more compelling evidence obtained by starting with lower and more realistic magnetic field strengths will be possibly achieved in the future.

The literature on BNS mergers forming a BH-disk remnant is far more rich (see, e.g., Ref.~\refcite{Faber2012,Baiotti2017} for a review). In particular, different groups reported on GRMHD simulations of BNS mergers exploring the possibility of launching magnetically-driven jets and the potential connection with SGRB central engines (e.g., Ref.~\refcite{Rezzolla2011,Kiuchi:2014:41502,Dionysopoulou:2015:92,Kawamura:2016:064012,Ruiz2016}).
Compared to the NS-BH case, BNS mergers have at least two important advantages. First, as shown also via GRHD simulations (i.e.~without magnetic fields), disk masses $\gtrsim$\,0.1\,$M_\odot$ are easily obtained for a large part of the parameter space spanning the possible BNS masses and NS EOS. Second, very efficient magnetic field amplification mechanisms can operate during and after merger, leading most certainly to final magnetic field strengths in excess of $\sim$\,10$^{15}$\,G. One such mechanism is the Kelvin-Helmholtz (KH) instability developing in the shear layer separating the two NS cores right before merger (e.g., Ref.~\refcite{Rasio1999,Price2006,Kiuchi:2015:1509.09205}). Then, further amplification is likely to occur inside the remnant during the HMNS phase, via magnetic winding and the MRI (e.g., Ref.~\refcite{Siegel:2013:121302}). Finally, efficient MRI amplification may continue inside the accretion disk after the remnant has collapsed to a BH (the only amplification allowed in the NS-BH case). Both the KH instability and the MRI operate on scales that are too small to be fully resolved in current simulations and thus the attainable amplification factors can be much smaller, unless an effective subgrid approach is adopted (e.g., Ref.~\refcite{Palenzuela2015,Zrake2013,GiacomazzoSub2015,Radice2017}). 

To date, a number of simulations showed the emergence of ordered magnetic field structures aligned with the BH spin axis, but not the formation of a Poynting-flux dominated outflow (e.g., Ref.~\refcite{Rezzolla2011,Kiuchi:2014:41502,Dionysopoulou:2015:92,Kawamura:2016:064012}). 
In Ref.~\refcite{Ruiz2016}, however, an incipient jet was reported (see Fig.\,\ref{fig1}). Similarly to the analogous result obtained in Ref.~\refcite{Paschalidis2015} for NS-BH mergers, a mildly relativistic outflow was produced along the axis and within a magnetically-dominated funnel. The time of emergence is again of order $\sim$\,100\,ms and corresponds to the time necessary to increase the magnetic-to-fluid energy density ratio up to $B^2/8\pi\rho c^2$\,$\gtrsim$\,100. In this case, even starting with magnetic fields entirely confined inside the two NSs led to the formation of an incipient jet (slightly less powerful than in the case with initial fields extending also outside the NSs). One main difference with the other BNS merger simulations is again the initial magnetic field strength imposed, in excess of $\sim$\,10$^{16}$\,G. All other simulations started with a much lower field strengths, closer to what expected from observations (i.e.~$\sim$\,10$^{12}$\,$-$\,$10^{13}$\,G in the NS interior). Moreover, the post-merger evolution was not as long. Hence, 
other groups should in principle be able to reproduce this result by running longer simulations and with much stronger initial magnetic fields. At present, this confirmation is still missing. 

\begin{figure}[tpb]
\centering
  \includegraphics[width=0.485 \textwidth]{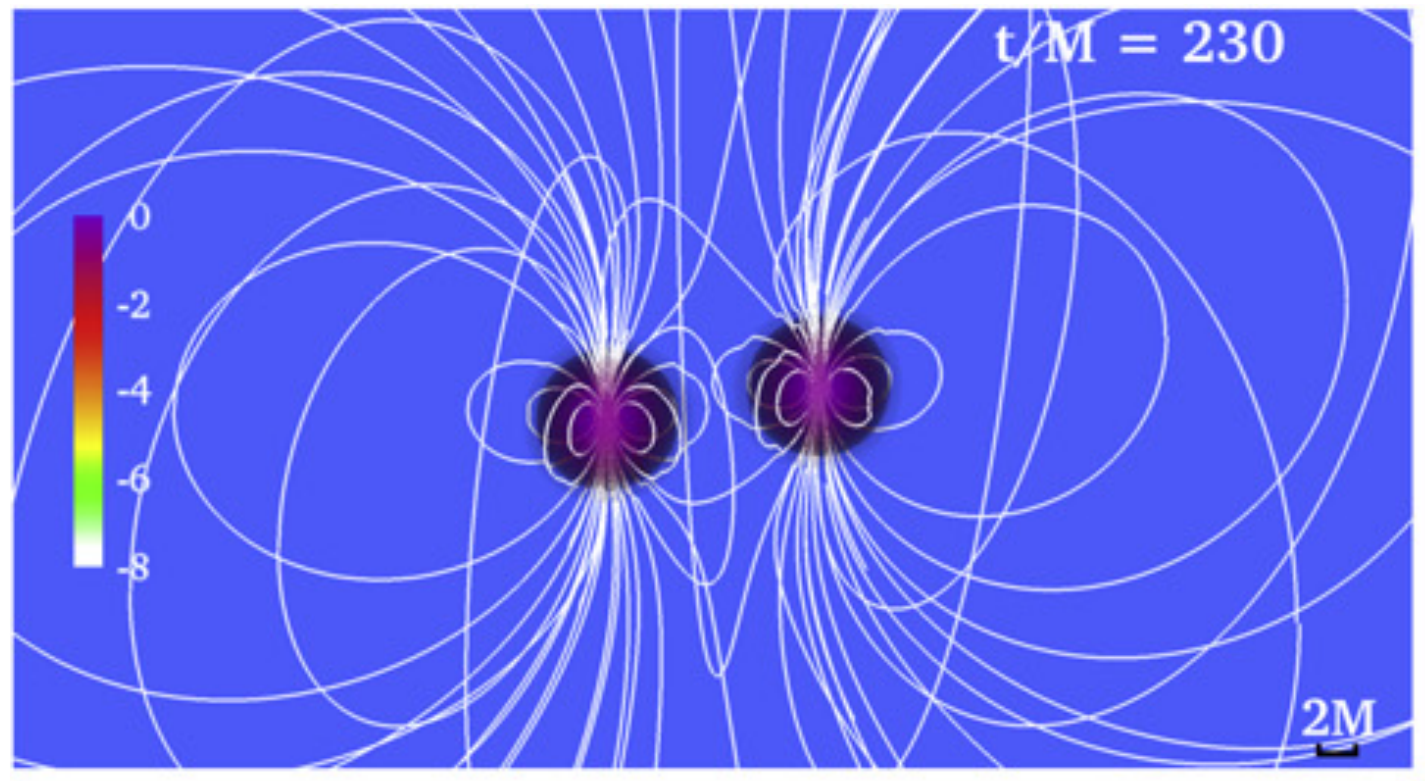}
  \includegraphics[width=0.485 \textwidth]{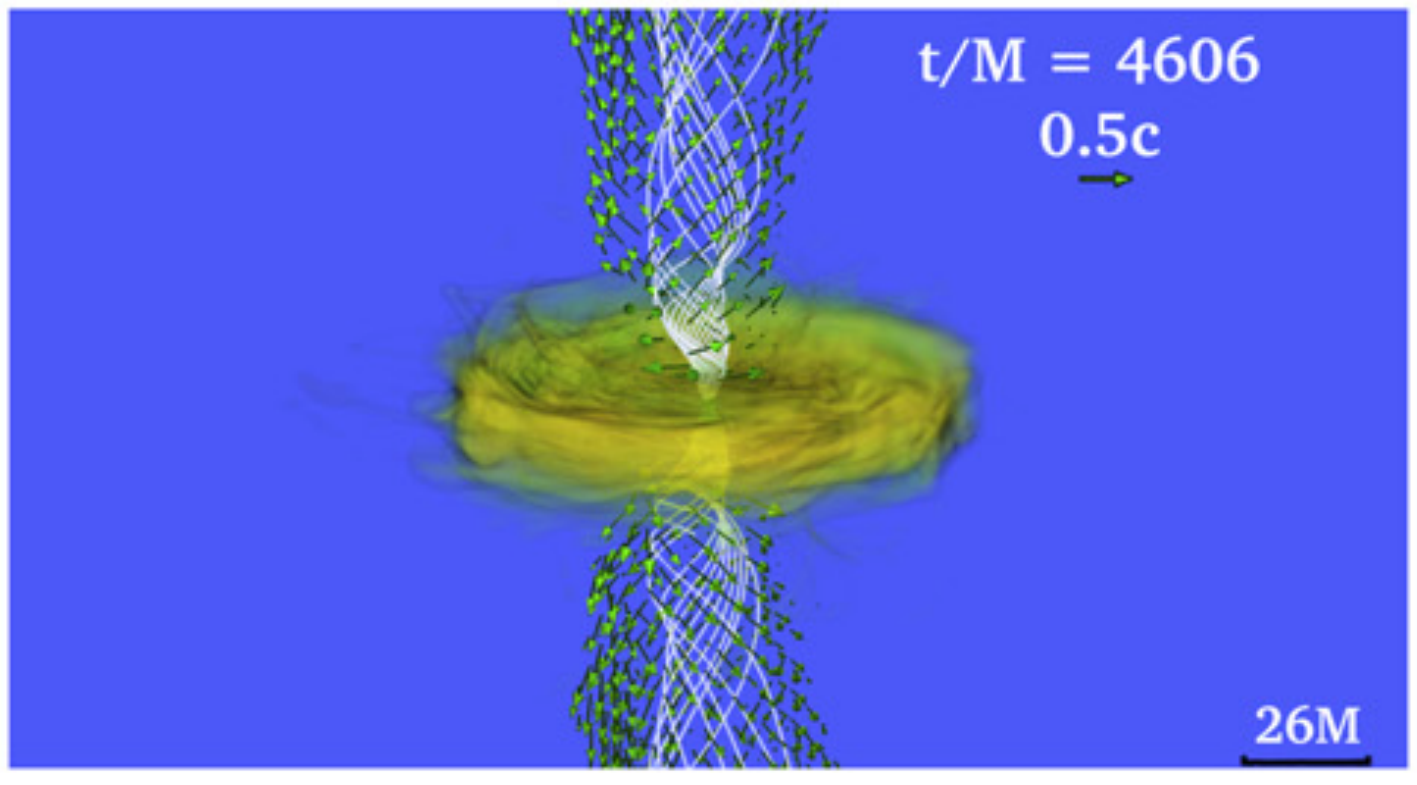}
\vspace*{8pt}
\caption{General relativistic magnetohydrodynamics simulation of a binary neutron star merger leading to the formation of an incipient jet (adapted from Ruiz et al. 2016). Rest-mass density is color coded and magnetic field lines are indicated in white. The left panel shows the two neutron stars before merger. 
The right panel shows the black hole-accretion disk system produced after merger and the ordered twister-like magnetic field structure formed along the black hole spin axis.}
\label{fig1}
\end{figure}
The above numerical results support the BH-disk scenario for SGRBs and favour a magnetically-driven jet over a neutrino-driven one. Nevertheless, merger simulations are still unable to provide a definite proof that for realistic initial conditions such a system can produce an incipient jet with the required properties.
Higher resolutions and longer simulations will be necessary for further progress, together with the inclusion of a more accurate description of the key physical ingredients (e.g., composition-dependent finite-temprature nuclear physics EOS, neutrino transport, and magnetic field evolution in both force-free and non-ideal MHD regimes). 


\subsection{Magnetar scenario}
\label{magnetar}

As discussed earlier in Section~\ref{engine}, BNS mergers can produce metastable or even stable massive NS remnants. 
If the remnant is a HMNS, it will collapse to a BH after a rather short lifetime $\lesssim$\,100\,ms and according to the scenario described in the previous Section~(\ref{BH}) the resulting BH-disk system could act as the central engine powering a SGRB jet. 
We now turn to consider the case in which the remnant is instead a long-lived NS.\footnote{Hereafter, we restrict to BNS mergers only, leaving aside NS-BH mergers.} This outcome might follow a significant fraction of all BNS mergers, posing the natural question on whether a SGRB jet could be produced in such circumstances. 
In particular, in the present Section we discuss the possibility that an incipient jet is launched shortly after merger by the long-lived NS itself, independently of its stable or metastable nature (i.e.~prior to a possible collapse to a BH). We shall refer to this scenario as the ``magnetar'' scenario. 

Indications that at least some SGRBs might be produced by a NS remnant rather than a BH-disk system come from specific features in the soft X-ray emission following the burst, detected by the {\it Swift} satellite\cite{Gehrels2004-Swift} in a significant fraction of events. 
These features are possibly indicative of ongoing central engine activity on timescales that extend way beyond the typical accretion timescale of a disk onto a stellar-mass BH ($\lesssim$\,1\,s) and include the so-called ``extended emission''\cite{Norris2006,Gompertz2013}  lasting $\sim$\,10\,$-$\,100\,s, X-ray flares\cite{Barthelmy2005b,Campana2006}, and X-ray ``plateaus''\cite{Rowlinson2010,Rowlinson2013,Lue2015}, in some cases appearing together in the same event (e.g., GRB 050724). 
In particular, the X-ray plateaus represent the most serious challenge for BH-disk central engines. They can last for minutes to hours (full range is $\sim$\,10$^{2}$\,$-$\,10$^{5}$\,s; e.g., Ref.~\refcite{Rowlinson2013}), maintaining a rather shallow decay that is inconsistent with the steeper lightcurves characterizing jet afterglows and suggesting instead persistent energy injection from a longer-lived NS remnant\footnote{Even the presence of fallback accretion, while offering a viable explanation for an X-ray flare\cite{Rosswog2007}, would hardly produce plateau-like emission.} 
(see, e.g., Ref.~\refcite{Kumar2015} and Refs.~therein).

The above considerations revived the idea that the SGRB central engine could be a long-lived and strongly magnetized NS (e.g., Ref.~\refcite{Zhang2001,Gao2006,Metzger2008}), naturally able to produce an X-ray plateau via electromagnetic spindown radiation.
In this case, after the jet has been launched there is still a very large energy reservoir given by the NS rotational energy ($\sim$\,10$^{52}$\,erg) that can be extracted via the magnetic field. This would also easily explain the large amount of energy associated with the X-ray plateaus, which can even be as large as the energy of the prompt gamma-ray emission. 

Modelling of X-ray plateaus as spindown radiation from a rotating magnetized NS was attempted (e.g., Ref.~\refcite{Rowlinson2013,Rowlinson2014,Lue2015}) assuming that the emission is regulated by the simple dipole spindown formula $L_\mathrm{X}$\,$=$\,$L_0 (1+t/t_\mathrm{sd})^{-2}$, where
\begin{equation}
L_0\sim10^{49} (B_\mathrm{pole}/10^{15}\,\mathrm{G})^2 (P/\,\mathrm{ms})^{-4}~\mathrm{erg/s}\, ,
\label{spindown}
\end{equation}
\begin{equation}
t_\mathrm{sd}\sim3 \times10^{3} (B_\mathrm{pole}/10^{15}\,\mathrm{G})^{-2} (P/\,\mathrm{ms})^{2}~\mathrm{s}\, ,
\label{spindown}
\end{equation}
and $B_\mathrm{pole}$ and $P$ are the dipolar magnetic field strength at the pole and the initial rotation period, respectively.
By directly fitting the X-ray plateau data, a rather satisfactory match was found for a good fraction of events. In a limited number of cases, even the late-time decay of the lightcurves was found consistent with the characteristic $\propto$\,$t^{-2}$ profile. Some other lightcurves show instead a more abrupt decay, interpreted by different authors as the sign of a late time collapse to a BH (e.g., Ref.~\refcite{Rowlinson2010,Rowlinson2013,Lasky2014}).
Overall, the above fitting procedure revealed typical values of $B_\mathrm{pole}$\,$\sim$\,10$^{15}$\,G and $P$\,$\sim$\,few\,ms, consistent with a millisecond magnetar. 

One main limitation of the above approach is that it neglects the presence of a baryon polluted environment surrounding the merger site. In a more realistic situation, the emerging signal is the result of a reprocessing of the spindown radiation across such an environment. 
Starting from the merger time, the system is likely to evolve as follows (e.g., Ref.~\refcite{Yu2013,Metzger2014b,Siegel:2016a,Siegel:2016b}; see Fig.\,\ref{fig2}).
At first, the long-lived (supramassive or stable) NS remnant is characterized by strong differential rotation, ongoing magnetic field amplification and neutrino cooling. During an early evolution phase, lasting $<$\,1\,s, neutrino and/or magnetically driven baryon-loaded winds are expelled isotropically causing substantial pollution (e.g., Ref.~\refcite{Dessart:2009:1681,Siegel:2014:6,Siegel2017}). This slowly expanding material ($v$\,$\lesssim$\,0.1\,c) can easily amount to 10$^{-3}$\,$-$\,10$^{-2}\,M_\odot$ and is additional to the earlier and faster dynamical ejecta associated with the merger process itself.
\begin{figure}[tpb]
\centering
  \includegraphics[width=0.99 \textwidth]{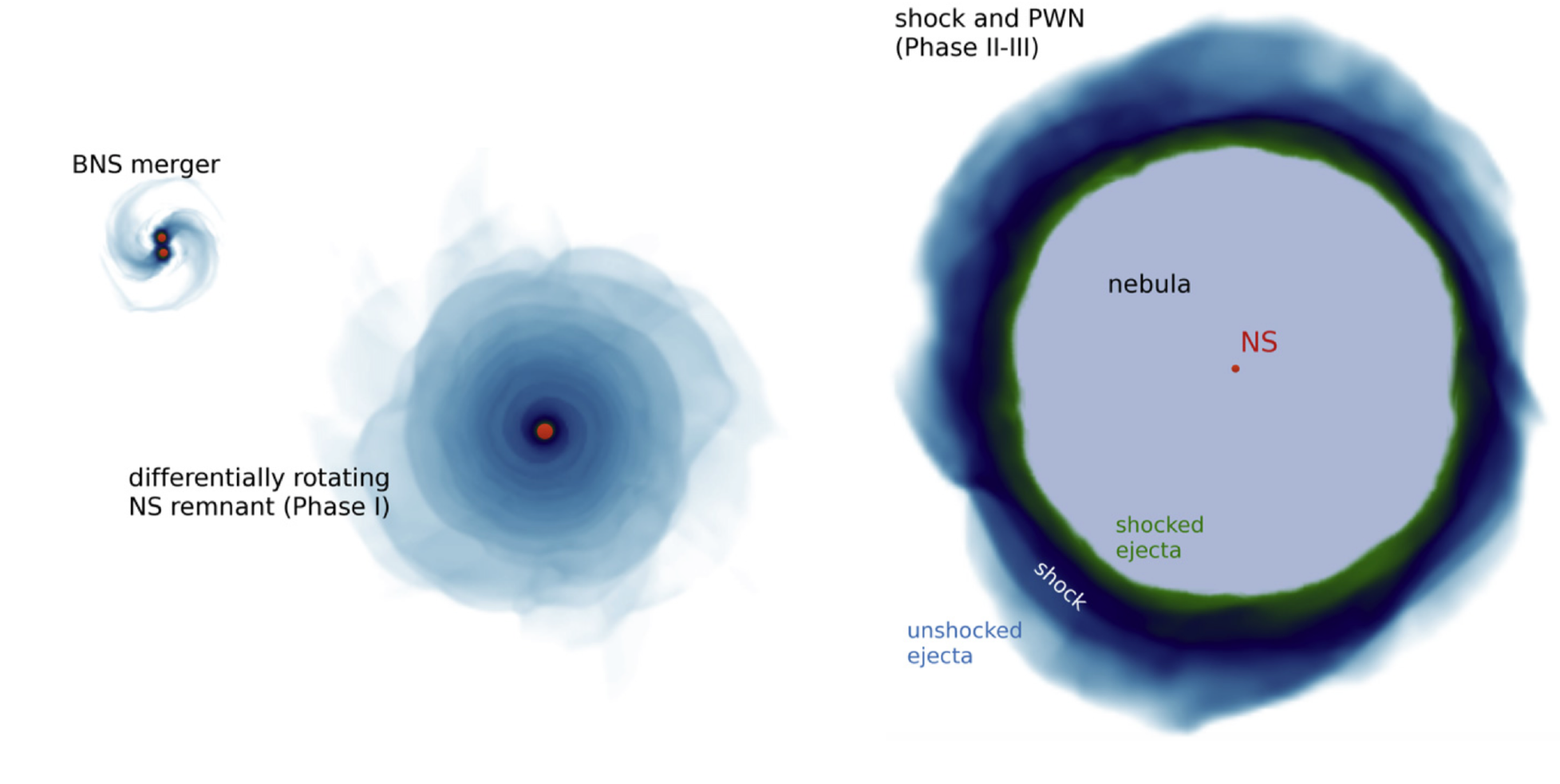}
\vspace*{8pt}
\caption{Evolution of a long-lived BNS merger remnant (adapted from Siegel \& Ciolfi 2016a). Left and Center: A BNS merger forms a hot and differentially rotating NS that ejects an isotropic baryon-loaded wind (Phase I). Right (on larger scale): Once wind ejection is suppressed, the cooled down NS starts emitting electromagnetic spindown radiation. The latter inflates a photon-pair plasma nebula (analogous to a pulsar wind nebula, PWN) that drives a shock sweeping up all the ejecta material into a thinner, hotter, and faster expanding shell (Phase II). Spindown emission from the NS continues while the nebula and the ejecta shell keep expanding (Phase III).
}
\label{fig2}
\end{figure}
As the NS cools down and magnetic field amplification is no longer active, the mass loss rate associated with baryon-loaded winds rapidly fades away resulting in a decreasing ambient density in the close surroundings of the NS. As density gets sufficiently low, the right conditions to build a magnetosphere are met and the strongly magnetized NS starts losing rotational energy via electromagnetic spindown radiation.
This sets the beginning of a new phase, in which spindown radiation inflates a photon-pair plasma nebula inside the shell of optically thick material previously ejected. Soon, the high pressure of the nebula drives a strong shock across the ejecta, heating up the material and further accelerating the expansion. Then, the evolution proceeds with the expansion of both the ejecta shell and the nebula, while the spindown of the NS continues. 
In the meanwhile, energy is partially lost by the system as photospheric emission from the outer ejecta layers, and this reprocessed radiation is what should power the observed X-ray plateau signals. A significant change in photometry and spectrum may also occur when the ejecta become optically thin and non-thermal radiation from the nebula can directly escape.
Moreover, if the massive NS is metastable (SMNS), the above evolution will be altered by the eventual collapse to a BH, switching off the spindown radiation and possibly producing a sudden energy release associated with the collapse itself. 

Models based on the above evolution phases were presented by different authors, enforcing spherical symmetry and adopting various simplifying assumptions on both the hydrodynamical evolution (obtained with a semi-analytical approach) and the radiative processes (e.g., Ref.~\refcite{Yu2013,Metzger2014b,Siegel:2016a,Siegel:2016b}). 
Depending primarily on the ejecta mass, the spindown luminosity, and the assumed ejecta opacity, the reprocessed spindown-powered radiation was found to peak between the soft X-ray and the optical band. 
In particular, it was shown that X-ray transients can be produced by such a system with energies, luminosities, and durations broadly consistent with the full range of observed X-ray plateaus in SGRBs (e.g., Ref.~\refcite{Siegel:2016a,Siegel:2016b}).
These results confirm the viability of an electromagnetic spindown origin for the X-ray plateaus, even though they are accompanied by a number of caveats to be resolved with various improvements (actual hydrodynamical simulations in special relativity, dropping the spherical symmetry, better treatment of the radiative processes, and so on). 
Moreover, a systematic comparison with data has not yet been attempted and a satisfactory explanation of various photometric and spectral features is still missing (e.g., late time decay of the lightcurves, non-thermal spectral components). 

As a side note, these luminous and highly isotropic spindown-powered signals also represent promising counterparts to the GW emission from a BNS merger, independently from a SGRB detection. They could even be produced in absence of a successful jet.\footnote{For instance, it was proposed that some of the softest X-ray Flashes (XRFs), which are commonly considered a sub-class of long GRBs, could have instead a BNS merger origin and be powered by the spindown radiation from a long-lived NS remnant (Ref.~\refcite{Ciolfi2016}; see also Ref.~\refcite{Metzger2008}).} 
Detecting these signals in the soft X-ray band after a GW trigger is currently a real challenge, due to the very short (minutes to hours) duration in combination with the small fields of view of present detectors in this band (e.g., {\it Swift} XRT). Even in the follow-up campaign of GW170817, the source was not properly covered by soft X-ray telescopes for several hours after merger (see Section~\ref{GRB170817}).
Future missions with large field of view detectors at $\sim$\,keV energies (e.g., Ref.~\refcite{THESEUS-WP,THESEUS-MM}) will offer the opportunity to catch systematically this early emission. 

If we admit the possibility that long-lived NS remnants are responsible for SGRBs with X-ray plateaus, there are some important consequences. First, we have to exclude a NS-BH origin for at least those SGRBs.  
Second, the fraction of SGRBs having an X-ray plateau ($\sim$\,50\% according to Ref.~\refcite{Rowlinson2013}) gives, among all BNS mergers leading to a SGRB, the minimum fraction having a long-lived NS remnant.\footnote{This fraction can be higher, e.g., if some of the SGRBs with no X-ray plateau have a NS-BH origin and/or if some of the long-lived NS remnants do not produce a detectable X-ray plateau because the emission peaks outside the soft X-ray band).} 
Finally and most importantly, the same system has to be able to launch a jet and produce the prompt SGRB emission. In this case, however, much less is known on the possible mechanisms that could make it feasible. 
BNS merger simulations show that a long-lived NS remnant can have a torus-shaped outer envelope which might persist for a timescale similar or even longer than the typical accretion timescale of a disk surrounding a stellar-mass BH (see Section \ref{TR} and Fig.\,\ref{fig4}). 
Nevertheless, this is not an accreting system analogous to a BH-disk and therefore it is unclear if an accretion-powered jet would be viable. 
Another main difficulty is baryon pollution. 
In the previous Section~\ref{BH}, we discussed how a BNS merger leading to a short-lived HMNS and then to a BH is likely to pollute the environment at levels that are prohibitive for a successful neutrino-powered jet. If the remnant is a long-lived NS, the level of baryon pollution can be orders of magnitude higher in density (in particular along the spin axis; e.g., Ref.~\refcite{Ciolfi2017}) and thus the conditions are even less favourable, despite the fact that the total energy emitted in neutrinos could be a factor of a few higher (e.g., Ref.~\refcite{Perego2017}).

Magnetically-powered jets are again the most promising solution. Even if no Blandford-Znajek mechanism is available in this case, the large rotational energy of the long-lived NS (easily exceeding $10^{52}$\,erg), if efficiently channeled via the magnetic field, would in principle be sufficient to power a SGRB jet. 
Nevertheless, how this would be realized is unclear. 
GRMHD simulations of differentially rotating NSs with an imposed monotonically decreasing ``j-constant'' rotation profile found that when the star is endowed with an ordered poloidal magnetic field aligned with the spin axis a collimated Poynting-flux dominated outflow is naturally generated\cite{Duez2006b,Kiuchi:2012:86,Siegel:2014:6}. However, this outcome is very sensitive to the chosen magnetic field geometry\cite{Siegel:2014:6} and there is no guarantee that an actual merger would be able to build the necessary magnetic field structure within the required short time window. 
A direct way to test this possibility is via GRMHD simulations of BNS mergers leading to a long-lived NS remnant. Very few simulations of this type have been performed so far (e.g., Ref.~\refcite{GiacomazzoPerna,Palenzuela2015,Endrizzi:2016:164001,Ciolfi2017}), showing no indication of jet formation.  
In particular, Ref.~\refcite{Ciolfi2017} studied the magnetic field structure up to $\sim$\,50\,ms after merger and found an emerging twister-like structure aligned with the spin axis, but no sign of a collimated outflow.
As suggested by the recent results obtained within the BH-disk scenario by Ref.~\refcite{Ruiz2016} (see Section~\ref{BH}), future simulations with much longer post-merger evolution and stronger initial magnetic fields (possibly extending also outside the two NSs) will provide a more stringent test. 


\subsection{Time-reversal scenario}
\label{TR}

The production of a jet in the magnetar scenario is highly uncertain, in part because of the unknown launching mechanism, in part due to the very high levels of baryon pollution which could obstruct the propagation of a collimated outflow. The BH-disk scenario, on the other hand, is challenged by the presence of X-ray plateaus in a fraction of SGRBs, suggesting energy injection from a long-lived NS remnant.
This leads to a dichotomy in which none of the two scenarios appear fully satisfactory, at least for those SGRBs accompanied by an X-ray plateau. 

In order to overcome the problems of the two scenarios, an alternative ``time-reversal'' (TR) scenario was recently proposed\cite{Ciolfi:2015:36}. In this new framework, a SGRB with an X-ray plateau is the result of the following phenomenology (see Fig.\,\ref{fig3}).
\begin{figure}[tpb]
\centering
  \includegraphics[width=0.99 \textwidth]{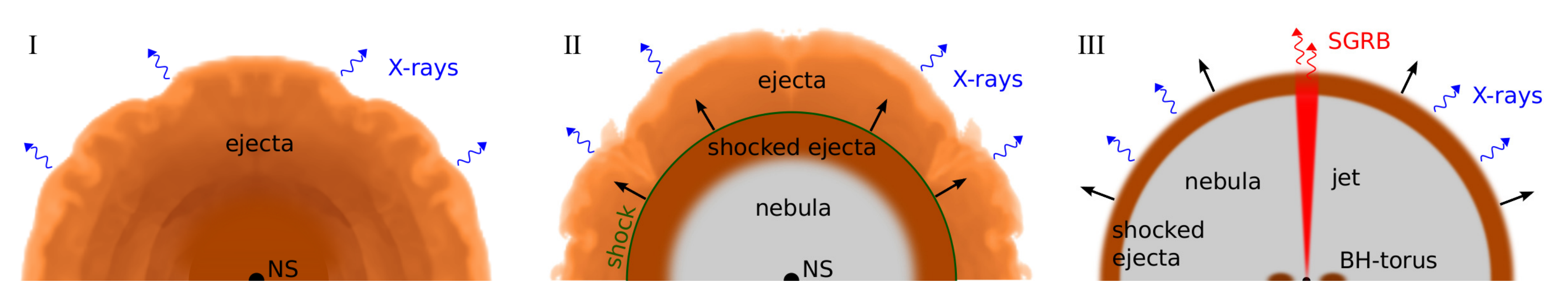}
\vspace*{8pt}
\caption{``Time-reversal'' evolution phases (from Ciolfi \& Siegel 2015): (I) the differentially rotating supramassive NS ejects a baryon-loaded and highly isotropic wind; (II) the cooled-down NS emits electromagnetic spindown radiation inflating a photon-pair plasma nebula that drives a shock through the ejecta; (III) the NS collapses to a BH launching a jet that drills through the nebula and the ejecta shell, eventually producing the prompt SGRB emission. Spindown radiation emitted by the NS before the collapse diffuses outward on much longer timescales and the corresponding signal persists even after the SGRB, possibly as an X-ray plateau.
}
\label{fig3}
\end{figure}
A BNS merger produces a long-lived but metastable NS remnant (i.e.~a supramassive NS) evolving as described in the previous Section (\ref{magnetar}) and in Fig.\,\ref{fig2}, with the formation of an expanding nebula surrounded by an optically thick ejecta shell and continuous energy injection form the NS via electromagnetic spindown radiation. 
However, since the NS is metastable, its centrifugal support will at some point become insufficient to prevent the collapse to a BH. As the scenario assumes, this leads to the formation of a BH-disk system able to launch a relativistic jet, in analogy with the standard BH-disk scenario (Section~\ref{BH}). 
This differs from the magnetar scenario, in which a stable or metastable long-lived NS launches the jet very shortly ($<$\,1\,s) after merger, before the onset of a spindown phase. In the TR case, the jet is launched only after the collapse to a BH. At that time, the close surroundings of the NS are no longer heavily baryon polluted and therefore an incipient jet finds no obstacle to its propagation. 
Once the jet is launched, it can easily drill through the nebula and the ejecta shell, finally breaking out of the system and producing the SGRB prompt emission.
Because of the high optical depth of both the nebula and the ejecta, the energy given off by NS via spindown radiation {\it before} the collapse remains stored for a long time before emerging from the outer ejecta layers and possibly producing a detectable X-ray plateau. As a result, while the jet propagation is essentially unaffected, the nearly isotropic spindown-powered signal is spread over much longer timescales and can still be observed for a long time even {\it after} the gamma-rays. 
In other words, the prompt SGRB emission and part of the spindown-powered emission (i.e.~the X-ray plateau) are observed in a reverse order with respect to the corresponding energy release from the central engine.

It is worth stressing that within the TR framework the observed duration of the X-ray plateaus is not an indication of the lifetime of the long-lived NS remnant, but rather of the diffusion timescale across the environment (nebula and ejecta) minus the light trave time. 
A simple calculation carried out in Ref.~\refcite{Ciolfi:2015:36} shows that spindown-powered emission lasting up to hours can be explained, thus covering the full range of observed X-ray plateau durations. 
Another interesting point suggested in Ref.~\refcite{Ciolfi:2015:36} is that the break-out of the strong shock initially produced by the spindown radiation (see Fig.\,\ref{fig3}) could produce a precursor transient detectable before the SGRB, with a time separation nearly corresponding to the duration of the spindown phase. This offers one possible explanation for SGRBs precursors observed $\sim$\,seconds (or more) before the prompt emission\cite{Troja2010}.  

The TR scenario offers a possible way to overcome the difficulties of the BH-disk and magnetar scenarios, by reconciling the putative needs for (i) an accreting BH and a baryon-free funnel to launch a jet and (ii) spindown radiation from a long-lived NS to power an X-ray plateau.\footnote{The idea of a time reversal was also adopted with a different phenomenology in Ref.~\refcite{Rezzolla2015}.}
Note that for very short remnant lifetimes $\lesssim$\,100\,ms (typical for HMNSs) there is no proper spindown phase and the ordinary BH-disk picture is recovered. 
In this respect, the TR model can be seen as an extension or modification of the BH-disk scenario that applies to SGRBs with an X-ray plateau, where the difference is that the energy powering the latter is not released by the remnant system during or after jet formation, but before the jet is launched.
Within this framework, we can easily explain why some SGRBs show an X-ray plateau and others do not. 
Moreover, as for the magnetar model, the fraction of SGRBs with a detected X-ray plateau gives the minimum fraction of all BNS mergers producing a SGRBs that have a long-lived NS remnant.

The main potential difficulty of this scenario is related to the assumption that a delayed collapse of the remnant would still form a BH surrounded by a massive ($\sim$\,0.1\,$M_\odot$) accretion disk, i.e.~a system able to launch a relativistic jet. 
As recently pointed out in Ref.~\refcite{Margalit2015}, the collapse of a uniformly rotating NS can hardly leave an accretion disk of significant mass, even if the NS is close to its maximal rotation.
This conclusion is based on the criterion that, given a uniformly rotating NS, all the material having a specific angular momentum smaller than the specific angular momentum at the ISCO of a Kerr BH with the same total mass and angular momentum will end up inside the horizon in case of collapse. 
Earlier numerical simulations of uniformly rotating NSs collapsing to a BH (e.g., Ref.~\refcite{Shibata2000,Baiotti:2005:24035}) also found negligible disk masses. 
The natural conclusion is that only a SMNS collapsing before it has settled to uniform rotation could lead to a viable BH-disk SGRB central engine. 
Therefore, the question on how a SMNS rearranges towards a uniformly rotating configuration and on which timescales becomes crucial. 

Current simulations of BNS mergers forming a long-lived NS remnant show that a few ms after merger the system reaches a nearly axisymmetric configuration which appears rather stationary on timescales of $\sim$\,10\,ms (e.g., Ref.~\refcite{Ciolfi2017}). At this stage, the angular velocity profile along the equatorial plane is characterized by a central core with a relatively slow rotation rate smoothly connected to an outer envelope with much higher angular velocity, increasing up to a typical radius of $\sim$\,10\,$-$\,20\,km and then decreasing for larger radii (see Fig.\,\ref{fig4}, left panel). The decreasing profile follows a nearly Keplerian behaviour, which indicates that the outer envelope is nearly orbiting, i.e.~the centrifugal support is dominant while pressure gradients provide a minor contribution.
Rotation profiles of this kind were first pointed out in Ref.~\refcite{Kastaun:2015:064027} and then confirmed in a number of studies (e.g., Ref.~\refcite{Endrizzi:2016:164001,Kastaun:2016,Kastaun2017,Hanauske2017,Ciolfi2017}).
\begin{figure}[tpb]
\centering
  \includegraphics[width=0.49 \textwidth]{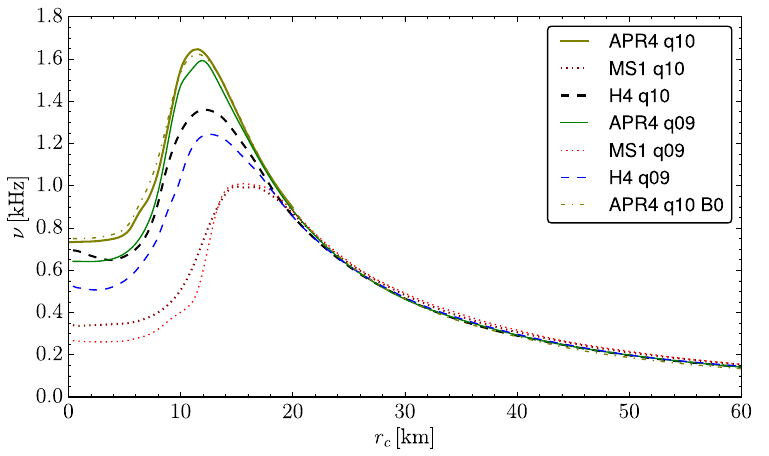}
  \includegraphics[width=0.495 \textwidth]{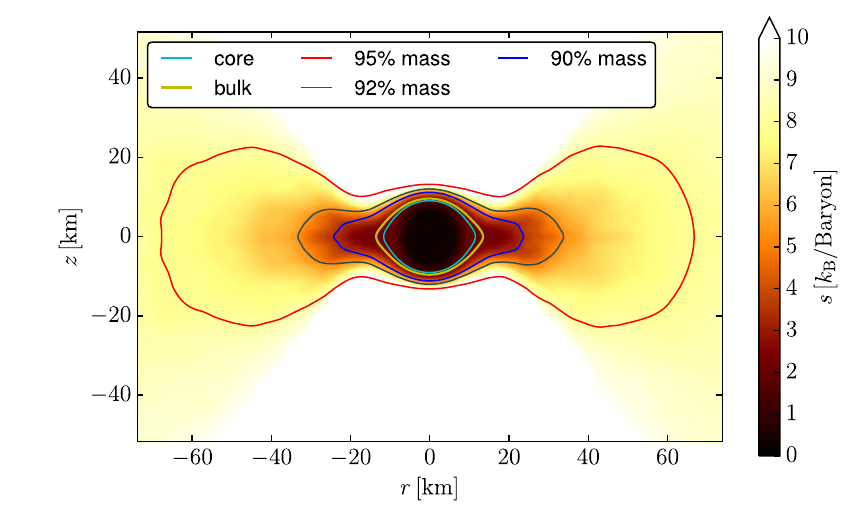}
\vspace*{8pt}
\caption{Left: Rotation profile of the NS remnant on the equatorial plane 20~ms after merger for various BNS merger models (from Ciolfi et al. 2017). Right: Meridional structure of a long-lived NS remnant 18~ms after merger (from Kastaun et al. 2016). 
}
\label{fig4}
\end{figure}
The meridional view of the system (see Fig.\,\ref{fig4}, right panel) reveals a configuration reminescent of a central object surrounded by a thick accretion disk. However, this is not an accreting system analogous to a BH surrounded by a disk. 
In this case, there is a single object composed by a slower spinning and spheroidal central core continuously connected with a faster spinning and lower density torus. 
Moreover, within the few tens of ms covered by the simulations, the flow of material close to the NS is mostly outgoing, also along the spin axis\cite{Ciolfi2017}. 

The long-term evolution of such a system and the associated timescales are unknown. 
However, a plausible evolutionary path should be as follows.
As the baryon-loaded winds are suppressed (within a fraction of a second, due to cooling and saturation of the magnetic field amplification) and the matter density in the polar regions drops, the emission of electromagnetic spindown radiation becomes possible, regulated by the dipolar field of the central core.
While the evolution proceeds, viscous effects associated with magnetic fields and turbulence in the torus result in a gradual matter transport towards the center, increasing the mass of the slowly rotating central core.
This can also transport some angular momentum inwards, partially balancing the losses associated with the spindown radiation. 
At the same time, the outer layers of the torus may expand outwards to compensate the angular momentum redistribution, effectively enlarging the torus itself (see, e.g., Ref.~\refcite{Pringle1981}).
By surviving long enough, the whole long-lived NS (central core plus torus) will eventually reduce to a uniformly rotating object, but the timescale for this to happen, $\tau_\mathrm{uniform}$, might be significantly longer than the timescale for removal of differential rotation in the central core alone ($\lesssim$\,0.1\,s). 
On the other hand, if the core collapses on timescales $\tau_\mathrm{collapse}< \tau_\mathrm{uniform}$, the outer torus will most likely survive the collapse and remain outside the horizon, forming an accretion disk. 
The question on the viability of the TR scenario can then be formulated in this way: for a given SGRB with an X-ray plateau, can the hypothetical long-lived NS survive long enough to release the energy powering the X-ray plateau (via spindown radiation), but also short enough to result in a BH surrounded by a massive ($\sim$\,0.1\,$M_\odot$) accretion disk?

The above considerations suggest a revision of some key aspects of the TR scenario as originally presented in Ref.~\refcite{Ciolfi:2015:36}. 
First, the timescale for the removal of differential rotation $t_\mathrm{dr}$ considered in Ref.~\refcite{Ciolfi:2015:36} ($\sim$\,100\,ms for magnetic field strengths of 10$^{15}$\,G; Ref.~\refcite{Shapiro2000,Siegel:2014:6}) applies to the central core, but the long-lived remnant is also composed by a faster spinning external torus which can persist on much longer timescales. In other words, at $t$\,$\sim$\,$t_\mathrm{dr}$ only the central core achieves a nearly uniform rotation, not the whole remnant. Concerning matter ejection via baryon-loaded winds, this should still be significantly suppressed on timescales $<$\,1\,s, as the remnant rapidly cools down and magnetic field amplification gets closer to a saturation. It remains unclear, however, if some residual baryon wind ejection persists on a longer timescale maintaining some level of baryon pollution in the vicinity of the remnant.  
Second, the electromagnetic spindown emission of the remnant does not correspond to that of a uniform spheroidal rotator with constant moment of inertia and the angular momentum evolution is altered by the ongoing rearrangement of the system. This could make the classic dipole spindown formula inadequate. 
Moreover, as pointed out above, the baryon content in the magnetosphere might be non negligible. 
Third, the idea that the scenario could work for lifetimes as long as canonical spindown timescales (minutes to hours for magnetar-like field strengths) is unlikely. 
The longest remnant lifetime which could still produce a massive disk after collapse is probably much shorter, i.e.~no longer than tens of seconds or even less. 
Note, however, that for a remnant lifetime of only 1\,$-$\,10\,s realistic dipolar magnetic field strengths of up to a few 10$^{15}$\,G should still be sufficient to explain the energy budget of the observed X-ray plateaus.\footnote{For instance, if we consider a rather common X-ray plateau energy of 10$^{49}$\,erg\,\cite{Rowlinson2010,Rowlinson2013,Lue2015} and an efficiency of $\sim$\,0.1 for the conversion of emitted spindown energy into observable X-ray signal, a spinning down NS remnant with initial central rotation rate of 1\,ms and a dipolar magnetic field of 10$^{15}$\,G ($3\times10^{15}$\,G) would take about 10\,s (1\,s), according to the canonical spindown formula, to release the necessary amount of energy (i.e.~10$^{50}$\,erg). }
For clarity, we stress again that the plateau duration is regulated by the optical depth of the environment at large scales (see Fig.\,\ref{fig2}) and the consequent diffusion timescales allow the escaping spindown-powered signal to last much longer than the lifetime of the remnant itself.

With the above revisions, the TR scenario and in particular the idea that spindown emission prior to collapse and jet formation can be responsible for the observed X-ray plateaus remain a possible solution. Future theoretical studies and simulations will shed light on the evolution of the system beyond the $\sim$\,100\,ms timescale and provide stronger indications on the viability of the TR picture. 
On the observational side, a distinctive signature of the TR phenomenology that allows for a conclusive test is the relatively long time window between the maximum GW emission (corresponding to the time of merger) and the onset of the gamma-ray signal (SGRB prompt emission). 
For a HMNS collapsing to a BH within $\sim$\,100\,ms, the overall time window (considering the time to launch the jet and produce the gamma-ray signal) should be $<$\,1\,s . In the TR scenario, a typical SGRB with X-ray plateau should correspond to a remnant lifetime at least a factor of 10\,$-$\,100 longer, resulting in a time separation $>$\,1\,$-$\,10\,s. 
Therefore, future observations measuring both the time separation and the energy in the X-ray plateau will offer a way to validate or discard the model. 


\section{GW170817 and GRB 170817A}
\label{GRB170817}

We now focus the attention on the BNS merger event GW170817 and the associated gamma-ray signal GRB 170817A.
The present Section is not intended to provide an exhaustive summary of the large body of literature on this event, but only to discuss its compatibility with the scenarios presented in Section~\ref{engine} depending on the nature of GRB 170817A (i.e.~a canonical SGRB vs.~a different type of GRB). 

The GW data, in combination with the electromagnetic counterparts observed across the entire spectrum, leave almost no doubt that GW170817 was the merger of two NSs\cite{LVC-BNS,LVC-MMA}. 
As for GRB 170817A, detected $\approx$\,1.7\,s after the peak of the GW signal (corresponding to the time of merger), the duration and spectrum of the gamma-ray emission are (marginally) consistent with the short-hard class of GRBs\cite{LVC-GRB,Goldstein2017}, allowing for a possible (and very tempting) interpretation of this event as the smoking gun evidence for the connection between SGRBs and BNS mergers. However, some elements cast doubts as to whether this merger was accompanied by a canonical SGRB. 
Taking into account the distance to the host galaxy NGC\,4993 ($\sim$\,40\,Mpc\,\cite{Hjorth2017,Im2017}) and thus the actual energy and luminosity of the burst, the latter resulted orders of magnitude less energetic than any observed SGRB with known redshift\cite{LVC-GRB} ($E_\mathrm{iso}\simeq3 \times 10^{46}$\,erg, $L_\mathrm{iso}\simeq2 \times 10^{47}$\,erg/s). 
Moreover, X-ray and radio afterglows emerging only around 9 and 16 days after merger\cite{Troja2017,Margutti2017,Haggard2017,Hallinan2017} were found inconsistent with an ultrarelativistic outflow pointing towards us.
Further modelling of the prompt and afterglow emission led to the conclusion that the observed gamma-ray signal was actually produced by a mildly relativistic outflow along line of sight (e.g., Ref.~\refcite{Mooley2018,Lazzati2018,Margutti2018,DAvanzo2018,Salafia2018}), which allows for different interpretations. In particular, we consider here the following alternatives:
\begin{enumerate}
\item The event produced a canonical SGRB jet pointing $\sim$\,20$^{\circ}$\,$-$\,40$^{\circ}$ away from us\footnote{
Note that GW data, combined with the known distance of NGC\,4993, only give upper limits to the viewing angle with respect to the orbital axis. Depending on the value assumed for the Hubble constant, the estimates give $\leqslant$\,36$^{\circ}$ or $\leqslant$\,28$^{\circ}$.\cite{LVC-GRB,LVC-Hubble}
} 
(e.g., Ref.~\refcite{Lazzati2018}). No emission from the jet core was observed, while the sub-energetic prompt gamma-ray emission was produced by a mildly relativistic and wide angle cocoon formed around the jet by the interaction of the latter with the baryon-polluted environment surrounding the merger site. 
The jet core contribution to the afterglow will eventually become observable along our direction in the future, causing a further rising of the afterglow lightcurves;
\item The incipient jet or outflow launched by the merger remnant was not powerful enough to successfully pierce through the baryon-polluted surroundings. In other words, the jet was choked (e.g., Ref.~\refcite{Mooley2018}). This resulted in a jet-less, wide angle (or nearly isotropic), and mildly relativistic outflow. GRB 170817A was not a canonical SGRB. 
\end{enumerate}
The two hypotheses predict a different future evolution of the multiwavelength afterglow lightcurves, although it is not clear whether future observations will be able to distinguish and, in that case, it might still take several months before a final answer is attainable\cite{Margutti2018,Nakar2018}.
For the purposes of the present work, it is however interesting to elaborate on both possibilities. 


\subsection{Canonical short gamma-ray burst}
\label{canonical}

Under the assumption that GRB 170817A was produced by a canonical SGRB jet oriented $\sim$\,30$^{\circ}$ away from us, the most popular SGRB scenario based on a BH-disk central engine (Section~\ref{BH}) is broadly compatible with the observations. 
This includes also the $\approx$\,1.7\,s time interval between the peak of the GW signal corresponding to the merger and the onset of the GRB. 
If this was an on-axis SGRB ($\Gamma$\,$\gtrsim$\,100), such a time interval would be largely dominated by the time for the remnant to collapse to a BH and launch the jet, implying a rather long survival time of the remnant ($\gtrsim$\,1\,s).
Nevertheless, the gamma-ray signal we observed was produced by a mildly relativistic ($\Gamma$\,$\sim$\,few) outflow moving along the line of sight, for which a significant part of the observed time delay is likely due to the outflow acceleration and propagation before becoming transparent and releasing the gamma-ray photons.\footnote{
From a simple calculation of the photospheric radius based on the fireball model and the formula $R_\mathrm{ph}$\,$\sim$\,$5\times10^{12} (L_\mathrm{iso}/10^{52}\,\mathrm{erg/s})(\Gamma/100)^{-3}$\,cm\,\cite{Kumar2015} and assuming that the outflow proceeds at constant $\Gamma$ since the beginning, we find that a delay of $1.7$\,s could be entirely explained by the propagation until transparency for $\Gamma$\,$\approx$\,4 (here $L_\mathrm{iso}=2\times10^{47}$\,erg/s).
}
We then conclude that the lifetime of the remnant, although maintaining an upper limit of $\sim$\,1\,s, might have been also much shorter.  
In particular, the possibility of a HMNS remnant collapsing to a BH in $\lesssim$\,0.1\,s is not in tension with the observations.\footnote{
Note that a HMNS remnant is also favoured by other constraints (not discussed here) based, e.g., on the GW signal and/or the optical/IR kilonova (e.g., Ref.~\refcite{Margalit2017,Shibata2017,Radice2018}, among others).}

If there was an electromagnetic spindown emission phase prior to collapse, most likley it lasted for a rather short time ($<$1\,s) and therefore, even assuming high magnetic field strengths 
$B_\mathrm{pole}$\,$\gtrsim$\,10$^{15}$\,G, an X-ray plateau of energy comparable to what typically observed in (a fraction of) SGRBs is not expected. 
This should then represent one of those cases in which the TR scenario reduces to the standard BH-disk scenario (see Section~\ref{TR}). 

The magnetar scenario (Section~\ref{magnetar}) cannot be completely excluded, although it seems much less favoured.
The major limitation remains the baryon pollution problem. This merger event likely produced, among the different ejecta components, a very massive ($\mathrm{few}\times10^{-2}\,M_\odot$) and slowly expanding ($v$\,$\lesssim$\,0.1\,$c$) baryon-loaded wind ejected isotropically in the early post-merger phase (e.g., Ref.~\refcite{Cowperthwaite2017,Kasen2017,Kasliwal2017,Pian2017,Tanvir2017,Smartt2017}, among others). 
In absence of a baryon-free funnel along the spin axis like the one that can only be created close to a BH, even a small fraction of this wind material expelled before jet formation would have easily obstructed its propagation.

Early observations in the soft X-ray band pointed on the source would have provided very helpful additional information. Unfortunately, such observations are missing. The most relevant upper limits were given by {\it MAXI} about 5 hours after the trigger ($L_X\lesssim10^{45}$\,erg/s at 2\,$-$\,10\,keV; Ref.~\refcite{Sugita2017}) and by {\it Swift} and {\it NuSTAR} respectively 15 and 17 hours after the trigger ($L_X\lesssim5\times10^{40}$\,erg/s at 0.3\,$-$\,10\,keV and $L_X\lesssim5\times10^{39}$\,erg/s at 
3\,$-$\,10\,keV, respectively; Ref.~\refcite{Evans2017}). However, due to their delay, the above observations would have missed most of the known SGRB X-ray plateaus. 
 As already pointed out in Section~\ref{magnetar}, covering the time window from a few minutes up to a few hours in the soft X-ray band is indeed very challenging at present. 

In the following, we speculate on the implications of a detection or a non-detection of a typical X-ray plateau for the different SGRB scenarios (summarized in Table\,1). 
An X-ray plateau non-detection would have further favoured a BH central engine, since in this case the absence of a sustained and long-lasting X-ray emission is expected. Within the TR framework, the remnant lifetime (as deduced from the $1.7$\,s delay of the GRB, see above) was likely too short to produce a typical X-ray plateau and thus a non-detection would be consistent. 
At the same time, the magnetar scenario would be further disfavoured. If a long-lived NS survived after the jet was launched, its spindown emission would have powered the ejecta leaving an observable signature. Excluding a release of energy in soft X-rays within the first few hours, a significant amount of energy (corresponding to a fraction of the residual rotational energy of the NS remnant) should have ended up in the budget comprising the kinetic energy of the ejecta plus the energy radiated in the optical/IR band.
Nevertheless, the estimates obtained from the kilonova signal can hardly accomodate such a large spindown input\cite{Margalit2017}.

An X-ray plateau detection would have been more puzzling. In this case, the BH-disk scenario would have to face the difficulty of producing a long-lasting signal with a central engine activity only lasting as long as the disk is not entirely accreted ($\sim$\,0.1\,$-$\,1\,s; see Section \ref{BH}). The TR scenario was proposed as a solution of this problem while maintaining the assumption that the SGRB jet is launched by a BH-disk system. However, it would be consistent with this event only if the NS remnant lifetime was sufficient to release via spindown radiation the energy required by the X-plateau. Given that the lifetime could not have been longer than $\sim$\,0.1\,$-$\,1\,s, this possibility is unlikely. 
Finally, the magnetar scenario would be more favourable, overcoming the energy budget limitations discussed above, but the baryon pollution problem would still represent an important element of doubt. 
In conclusion, no scenario would appear entirely satisfactory (see Table\,1).
\begin{table}[b]
    \begin{tabular}{ | l | p{3.9cm} | p{3.9cm} | }
    \hline
    early X-ray monitoring & plateau non-detection & plateau detection \\ \hline    \hline
    successful jet  & Most favourable case for {\bf BH-disk} scenario, broad agreement with expectations
     \vspace{0.2cm} \newline {\bf TR} scenario consistent for a short remnant lifetime of at most $\sim$\,0.1\,$-$\,1\,s (most likely case) \vspace{0.2cm} \newline {\bf Magnetar} scenario very much disfavoured, baryon pollution problem and possible tension with energy budget \vspace{0.4cm} & {\bf BH-disk} scenario challenged by a plateau-like emission \,\,lasting \,\,much longer than the accretion timescale  \vspace{0.2cm} \newline {\bf TR} scenario consistent for remnant lifetime $\gtrsim$\,1\,s (unlikely) \vspace{0.2cm} \newline {\bf Magnetar} scenario supported by X-ray plateau, but possibly hampered by baryon pollution \\ \hline
    choked jet & Same as successful jet for {\bf BH-disk} and {\bf TR} scenarios \vspace{0.2cm} \newline {\bf Magnetar} scenario in possible tension with the energy budget \vspace{0.2cm} & Same as successful jet for {\bf BH-disk} and {\bf TR} scenarios \vspace{0.2cm} \newline Most favourable case for {\bf Magnetar} scenario, {\color{white} \,\,.} energy budget problem mitigated \vspace{0.4cm} \\ \hline
    \end{tabular}
\vspace{0.6cm} \\ {\footnotesize Table~1. ~Implications of a successful vs.~choked jet for the central engine of GRB 170817A, assuming a detection or a non-detection of an X-ray plateau. \par}
\end{table}


\subsection{Choked jet}
\label{choked}

After considering GRB 170817A as a canonical SGRB observed off-axis, we now make the assumption that this GRB was instead produced by a failed or choked jet which was not able to break out of the baryon polluted merger surroundings. This implies that GRB 170817A belongs to a distinct type of GRBs and some of the arguments discussed in previous Sections that apply only to canonical SGRBs are no longer valid. 
Nevertheless, we can still explore the compatibility of this event with the different central engine scenarios under consideration. 

The main difference with the canonical SGRB hypothesis is that a choked jet is more favourable for the magnetar scenario. Indeed the prime difficulty of the latter, namely the production of a successful jet despite the very high levels of baryon pollution, is now mitigated.
In such circumstances, knowing the presence or absence of early X-ray emission (within minutes to hours from the merger time) and the associated energy would be of great aid to constrain the remnant lifetime and the central engine nature. 
Since a choked jet is conceivable also in the BH-disk and TR scenarios, our conclusions on those, with or without early X-ray emission, would still hold. 
On the other hand, the magnetar scenario could naturally explain the presence of a choked jet accompanied by long-lasting X-ray emission. The problem related to the energy budget (see Section~\ref{canonical}) would persist, but it would be mitigated. Without early X-ray emission, however, the energy budget would still represent a strong limitation, leaving serious doubts on the viability of a magnetar central engine. 

Summarizing (see also Table\,1), while a canonical SGRB would strongly favour a BH engine, in the choked jet case it is more difficult to guess which type of engine is more likely and early X-ray observations would have represented an even more crucial tool to discriminate. 


\section{Summary and Conclusions}
\label{conclusion}

Decades of investigation led to the notion that SGRBs are most likely associated with the merger of BNS and/or NS-BH binary systems. At the same time, the prompt emission of both long and short GRBs is commonly explained by assuming the presence of a highly relativistic and collimated outflow, i.e.~a jet. 
The combination of these two elements implies that the remnant of such mergers should be capable of launching a jet, at least under certain conditions. However, current models and numerical simulations are still unable to provide a final proof and, at present, the actual mechanism leading to jet formation and the nature of the central engine itself (BH vs. massive NS) remain uncertain.

In this Paper, we discussed such a critical aspect of the connection between SGRBs and BNS/NS-BH mergers. 
In particular, we considered three different scenarios for the SGRB central engine, pointing out the elements of strength and weakness of each scenario while taking into account the most recent observational constraints and theoretical results. These include the leading ``BH-disk'' scenario and the most discussed alternative, the ``magnetar'' scenario. The literature on SGRBs (and GRBs in general) refers almost entirely to one of these two frameworks. 
The third hypothesis we discussed is the recently proposed ``time-reversal'' scenario. 
Inspired by recent results obtained in BNS merger simulations, we proposed here a revision of the original time-reversal model, by which this scenario can still be considered a viable solution of the SGRB puzzle. 

In the last part of the Paper, we focussed the attention on the recent BNS merger event GW170817 and the accompanying gamma-ray signal GRB 170817A. At present, it is not clear whether this GRB was (i) a canonical SGRB pointing away from us or (ii) a different type of GRB in which the incipient jet was choked by the baryon polluted environment, resulting in a less collimated and mildly relativistic outflow. Currently, both hypotheses are consistent with the data at our disposal. A comparison with the predictions of the different scenarios led us to conclude that a BH-disk central engine is strongly favoured in the case of a successful jet (i.e.~a canonical SGRB), while the choked jet case leaves more doubts on the nature of the merger remnant (and thus more space for a long-lived NS central engine). 
We also discussed how an early observation of the source in the soft X-ray band (within the first few hours) would have provided key additional information to favour or disfavour the different scenarios.

The first GW and electromagnetic detection of a BNS merger opened the new field of multimessenger astronomy and astrophysics with GW sources. 
More and more BNS (and possibly NS-BH) merger events will be detected in the near future, likely offering new valuable insights into the origin of SGRBs and their progenitor systems.
Major progress towards a full understanding of the SGRB phenomenon, however, will also require a significant advancement of theoretical models and numerical simulations. 
In particular, in this Paper we encountered a number of key questions that cannot be directly answered only via new observational data.
Is the magnetic field the main driver of a SGRB jet? What is the role of neutrinos?
What makes different SGRB events have a different prompt gamma-ray emission as well as different soft X-ray features (i.e.~extended emission, X-ray plateaus, X-ray flares)?
Are there more than one SGRB central engine types?
How does a long-lived BNS merger remnant evolve on timescales longer than $\sim$\,100\,ms? Can a massive accretion disk be formed when the remnant collapses on timescales of 1\,$-$\,10\,s or longer?  
Which observational signatures would allow us to distinguish SGRBs with a BNS and a NS-BH origin? 
Only by addressing these (and other) urgent questions with more advanced models and simulations will allow us to fully exploit the scientific potential of future observations.


\section*{Acknowledgments}

We thank Wolfgang Kastaun, Daniel Siegel, and Om Sharan Salafia for many useful comments on the manuscript.


\bibliographystyle{ws-ijmpd_noeprint}

\begin{thebibliography}{137}

\bibitem{Kouveliotou1993}
C.~{Kouveliotou}, C.~A. {Meegan}, G.~J. {Fishman}, N.~P. {Bhat}, M.~S.
  {Briggs}, T.~M. {Koshut}, W.~S. {Paciesas} and G.~N. {Pendleton}, {\em
  Astrophys. J. Letters} {\bf 413} (August 1993) L101.

\bibitem{MacFadyen1999}
A.~I. {MacFadyen} and S.~E. {Woosley}, {\em Astrophys. J.} {\bf 524} (October
  1999) 262.

\bibitem{Bloom1998}
J.~S. {Bloom}, S.~G. {Djorgovski}, S.~R. {Kulkarni} and D.~A. {Frail}, {\em
  Astrophys. J. Letters} {\bf 507} (November 1998) L25.

\bibitem{Fruchter2006}
A.~S. {Fruchter}, A.~J. {Levan}, L.~{Strolger}, P.~M. {Vreeswijk}, S.~E.
  {Thorsett}, D.~{Bersier}, I.~{Burud}, J.~M. {Castro Cer{\'o}n}, A.~J.
  {Castro-Tirado}, C.~{Conselice}, et al., {\em Nature} {\bf 441} (May 2006) 463.

\bibitem{Galama1998}
T.~J. {Galama}, P.~M. {Vreeswijk}, J.~{van Paradijs}, C.~{Kouveliotou},
  T.~{Augusteijn}, H.~{B{\"o}hnhardt}, J.~P. {Brewer}, V.~{Doublier}, J.-F.
  {Gonzalez}, B.~{Leibundgut}, et al., {\em Nature} {\bf 395} (October 1998) 670.

\bibitem{Paczynski1986}
B.~{Paczynski}, {\em Astrophys. J. Letters} {\bf 308} (September 1986) L43.

\bibitem{Eichler:1989:126}
D.~{Eichler}, M.~{Livio}, T.~{Piran} and D.~N. {Schramm}, {\em Nature} {\bf
  340} (July 1989) 126.

\bibitem{Narayan1992}
R.~{Narayan}, B.~{Paczynski} and T.~{Piran}, {\em Astrophys. J. Letters} {\bf
  395} (August 1992) L83.

\bibitem{Berger2014}
E.~{Berger}, {\em ARA\&A} {\bf 52} (August 2014) 43.

\bibitem{Tanvir2013}
N.~R. {Tanvir}, A.~J. {Levan}, A.~S. {Fruchter}, J.~{Hjorth}, R.~A. {Hounsell},
  K.~{Wiersema} and R.~L. {Tunnicliffe}, {\em Nature} {\bf 500} (August 2013)
  547.

\bibitem{Berger2013}
E.~{Berger}, W.~{Fong} and R.~{Chornock}, {\em Astrophys. J. Letters} {\bf 774}
  (September 2013)   L23.

\bibitem{Lattimer1974}
J.~M. {Lattimer} and D.~N. {Schramm}, {\em Astrophys. J. Letters} {\bf 192}
  (September 1974) L145.

\bibitem{Lattimer1977}
J.~M. {Lattimer}, F.~{Mackie}, D.~G. {Ravenhall} and D.~N. {Schramm}, {\em
  Astrophys. J. Letters} {\bf 213} (April 1977) 225.

\bibitem{Li:1998:L59}
L.-X. {Li} and B.~{Paczy{\'n}ski}, {\em Astrophys. J.} {\bf 507} (November
  1998) L59.

\bibitem{Metzger2012}
B.~D. {Metzger} and E.~{Berger}, {\em Astrophys. J.} {\bf 746} (February 2012)
  ~48.

\bibitem{LVC-BNS}
B.~P. {Abbott}, R.~{Abbott}, T.~D. {Abbott}, F.~{Acernese}, K.~{Ackley},
  C.~{Adams}, T.~{Adams}, P.~{Addesso}, R.~X. {Adhikari}, V.~B. {Adya}, et al., 
  {\em Phys. Rev. Letters} {\bf 119} (October 2017)   161101.

\bibitem{LVC-MMA}
B.~P. {Abbott}, R.~{Abbott}, T.~D. {Abbott}, F.~{Acernese}, K.~{Ackley},
  C.~{Adams}, T.~{Adams}, P.~{Addesso}, R.~X. {Adhikari}, V.~B. {Adya}, et al., 
  {\em Astrophys. J. Letters} {\bf 848} (October 2017)   L12.

\bibitem{Cowperthwaite2017}
P.~S. {Cowperthwaite}, E.~{Berger}, V.~A. {Villar}, B.~D. {Metzger},
  M.~{Nicholl}, R.~{Chornock}, P.~K. {Blanchard}, W.~{Fong}, R.~{Margutti},
  M.~{Soares-Santos}, et al., {\em Astrophys. J. Letters} {\bf 848} (October 2017)   
  L17.

\bibitem{Kasen2017}
D.~{Kasen}, B.~{Metzger}, J.~{Barnes}, E.~{Quataert} and E.~{Ramirez-Ruiz},
  {\em Nature} {\bf 551} (November 2017) 80.

\bibitem{Kasliwal2017}
M.~M. {Kasliwal}, E.~{Nakar}, L.~P. {Singer}, D.~L. {Kaplan}, D.~O. {Cook},
  A.~{Van Sistine}, R.~M. {Lau}, C.~{Fremling}, O.~{Gottlieb}, J.~E. {Jencson}, 
  et al., {\em Science} {\bf 358} (December 2017) 1559.

\bibitem{Pian2017}
E.~{Pian}, P.~{D'Avanzo}, S.~{Benetti}, M.~{Branchesi}, E.~{Brocato},
  S.~{Campana}, E.~{Cappellaro}, S.~{Covino}, V.~{D'Elia}, J.~P.~U. {Fynbo}, 
  et al., {\em Nature} {\bf 551} (November 2017) 67.

\bibitem{Tanvir2017}
N.~R. {Tanvir}, A.~J. {Levan}, C.~{Gonz{\'a}lez-Fern{\'a}ndez}, O.~{Korobkin},
  I.~{Mandel}, S.~{Rosswog}, J.~{Hjorth}, P.~{D'Avanzo}, A.~S. {Fruchter},
  C.~L. {Fryer}, et al., {\em Astrophys. J. Letters} {\bf 848}
  (October 2017)   L27.

\bibitem{Smartt2017}
S.~J. {Smartt}, T.-W. {Chen}, A.~{Jerkstrand}, M.~{Coughlin}, E.~{Kankare},
  S.~A. {Sim}, M.~{Fraser}, C.~{Inserra}, K.~{Maguire}, K.~C. {Chambers}, et al., 
  {\em Nature} {\bf 551} (November 2017) 75.

\bibitem{LVC-GRB}
B.~P. {Abbott}, R.~{Abbott}, T.~D. {Abbott}, F.~{Acernese}, K.~{Ackley},
  C.~{Adams}, T.~{Adams}, P.~{Addesso}, R.~X. {Adhikari}, V.~B. {Adya}, et al., 
  {\em Astrophys. J. Letters} {\bf 848} (October 2017)   L13.

\bibitem{Goldstein2017}
A.~{Goldstein}, P.~{Veres}, E.~{Burns}, M.~S. {Briggs}, R.~{Hamburg},
  D.~{Kocevski}, C.~A. {Wilson-Hodge}, R.~D. {Preece}, S.~{Poolakkil}, O.~J.
  {Roberts}, et al., {\em Astrophys. J. Letters} {\bf 848}
  (October 2017)   L14.

\bibitem{Troja2017}
E.~{Troja}, L.~{Piro}, H.~{van Eerten}, R.~T. {Wollaeger}, M.~{Im}, O.~D.
  {Fox}, N.~R. {Butler}, S.~B. {Cenko}, T.~{Sakamoto}, C.~L. {Fryer}, et al., 
  {\em Nature} {\bf 551} (November 2017) 71.

\bibitem{Margutti2017}
R.~{Margutti}, E.~{Berger}, W.~{Fong}, C.~{Guidorzi}, K.~D. {Alexander}, B.~D.
  {Metzger}, P.~K. {Blanchard}, P.~S. {Cowperthwaite}, R.~{Chornock},
  T.~{Eftekhari}, et al., {\em Astrophys. J. Letters} {\bf 848} (October 2017)   L20.

\bibitem{Haggard2017}
D.~{Haggard}, M.~{Nynka}, J.~J. {Ruan}, V.~{Kalogera}, S.~B. {Cenko},
  P.~{Evans} and J.~A. {Kennea}, {\em Astrophys. J. Letters} {\bf 848} (October
  2017)   L25.

\bibitem{Hallinan2017}
G.~{Hallinan}, A.~{Corsi}, K.~P. {Mooley}, K.~{Hotokezaka}, E.~{Nakar}, M.~M.
  {Kasliwal}, D.~L. {Kaplan}, D.~A. {Frail}, S.~T. {Myers}, T.~{Murphy}, et al., 
  {\em Science} {\bf 358} (December 2017) 1579.

\bibitem{Mooley2018}
K.~P. {Mooley}, E.~{Nakar}, K.~{Hotokezaka}, G.~{Hallinan}, A.~{Corsi}, D.~A.
  {Frail}, A.~{Horesh}, T.~{Murphy}, E.~{Lenc}, D.~L. {Kaplan}, et al., 
  {\em Nature} {\bf 554} (February 2018) 207.

\bibitem{Lazzati2018}
D.~{Lazzati}, R.~{Perna}, B.~J. {Morsony}, D.~{L{\'o}pez-C{\'a}mara},
  M.~{Cantiello}, R.~{Ciolfi}, B.~{Giacomazzo} and J.~C. {Workman}, {\em ArXiv
  e-prints}  (December 2017).

\bibitem{Margutti2018}
R.~{Margutti}, K.~D. {Alexander}, X.~{Xie}, L.~{Sironi}, B.~D. {Metzger},
  A.~{Kathirgamaraju}, W.~{Fong}, P.~K. {Blanchard}, E.~{Berger},
  A.~{MacFadyen}, et al., {\em ArXiv e-prints}  (January 2018).

\bibitem{Nakar2018}
E.~{Nakar} and T.~{Piran}, {\em ArXiv e-prints}  (January 2018).

\bibitem{Rees1992}
M.~J. {Rees} and P.~{Meszaros}, {\em Mon. Not. R. Astron. Soc.} {\bf 258}
  (September 1992) 41P.

\bibitem{Rhoads1999}
J.~E. {Rhoads}, {\em Astrophys. J.} {\bf 525} (November 1999) 737.

\bibitem{Sari1999}
R.~{Sari}, T.~{Piran} and J.~P. {Halpern}, {\em Astrophys. J. Letters} {\bf
  519} (July 1999) L17.

\bibitem{Kumar2015}
P.~{Kumar} and B.~{Zhang}, {\em Phys. Rep.} {\bf 561} (February 2015) 1.

\bibitem{Rees1994}
M.~J. {Rees} and P.~{Meszaros}, {\em Astrophys. J. Letters} {\bf 430} (August
  1994) L93.

\bibitem{Rees2005}
M.~J. {Rees} and P.~{M{\'e}sz{\'a}ros}, {\em Astrophys. J.} {\bf 628} (August
  2005) 847.

\bibitem{Giannios2006}
D.~{Giannios}, {\em Astron. Astrophys.} {\bf 455} (August 2006) L5.

\bibitem{Lazzati2009}
D.~{Lazzati}, B.~J. {Morsony} and M.~C. {Begelman}, {\em Astrophys. J. Letters}
  {\bf 700} (July 2009) L47.

\bibitem{Paschalidis2017}
V.~{Paschalidis}, {\em Class. Quantum Grav.} {\bf 34} (April 2017)   084002.

\bibitem{Baiotti2017}
L.~{Baiotti} and L.~{Rezzolla}, {\em Rep. Prog. Phys.} {\bf 80} (September
  2017)   096901.

\bibitem{Kiuchi:2015:1509.09205}
K.~{Kiuchi}, P.~{Cerd{\'a}-Dur{\'a}n}, K.~{Kyutoku}, Y.~{Sekiguchi} and
  M.~{Shibata}, {\em Phys. Rev. D} {\bf 92} (December 2015)   124034.

\bibitem{Nagakura2014}
H.~{Nagakura}, K.~{Hotokezaka}, Y.~{Sekiguchi}, M.~{Shibata} and K.~{Ioka},
  {\em Astrophys. J. Letters} {\bf 784} (April 2014)   L28.

\bibitem{Murguia-Berthier2014}
A.~{Murguia-Berthier}, G.~{Montes}, E.~{Ramirez-Ruiz}, F.~{De Colle} and W.~H.
  {Lee}, {\em Astrophys. J. Letters} {\bf 788} (June 2014)  ~L8.

\bibitem{Murguia-Berthier2017a}
A.~{Murguia-Berthier}, E.~{Ramirez-Ruiz}, G.~{Montes}, F.~{De Colle},
  L.~{Rezzolla}, S.~{Rosswog}, K.~{Takami}, A.~{Perego} and W.~H. {Lee}, {\em
  Astrophys. J. Letters} {\bf 835} (February 2017)   L34.

\bibitem{Lazzati2017}
D.~{Lazzati}, D.~{L{\'o}pez-C{\'a}mara}, M.~{Cantiello}, B.~J. {Morsony},
  R.~{Perna} and J.~C. {Workman}, {\em Astrophys. J. Letters} {\bf 848}
  (October 2017)  ~L6.

\bibitem{Gottlieb2018}
O.~{Gottlieb}, E.~{Nakar} and T.~{Piran}, {\em Mon. Not. R. Astron. Soc.} {\bf
  473} (January 2018) 576.

\bibitem{Foucart2012}
F.~{Foucart}, {\em Phys. Rev. D} {\bf 86} (December 2012)   124007.

\bibitem{Pannarale2014}
F.~{Pannarale} and F.~{Ohme}, {\em Astrophys. J. Letters} {\bf 791} (August
  2014)  ~L7.

\bibitem{Demorest2010}
P.~B. {Demorest}, T.~{Pennucci}, S.~M. {Ransom}, M.~S.~E. {Roberts} and
  J.~W.~T. {Hessels}, {\em Nature} {\bf 467} (October 2010) 1081.

\bibitem{Antoniadis2013}
J.~{Antoniadis}, P.~C.~C. {Freire}, N.~{Wex}, T.~M. {Tauris}, R.~S. {Lynch},
  M.~H. {van Kerkwijk}, M.~{Kramer}, C.~{Bassa}, V.~S. {Dhillon}, T.~{Driebe}, 
  et al., {\em Science} {\bf 340} (April 2013)   448.

\bibitem{Belczynski2008}
K.~{Belczynski}, R.~{O'Shaughnessy}, V.~{Kalogera}, F.~{Rasio}, R.~E. {Taam}
  and T.~{Bulik}, {\em Astrophys. J. Letters} {\bf 680} (June 2008)   L129.

\bibitem{Lasota1996}
J.-P. {Lasota}, P.~{Haensel} and M.~A. {Abramowicz}, {\em Astrophys. J.} {\bf
  456} (January 1996)   300.

\bibitem{Piro2017}
A.~L. {Piro}, B.~{Giacomazzo} and R.~{Perna}, {\em Astrophys. J. Letters} {\bf
  844} (August 2017)   L19.

\bibitem{Gao2016}
H.~{Gao}, B.~{Zhang} and H.-J. {L{\"u}}, {\em Phys. Rev. D} {\bf 93} (February
  2016)   044065.

\bibitem{Mochkovitch1993}
R.~{Mochkovitch}, M.~{Hernanz}, J.~{Isern} and X.~{Martin}, {\em Nature} {\bf
  361} (January 1993) 236.

\bibitem{Blandford1977}
R.~D. {Blandford} and R.~L. {Znajek}, {\em MNRAS} {\bf 179} (May 1977) 433.

\bibitem{Ruffert:1999:573}
M.~Ruffert and H.-T. Janka, {\em Astron. Astrophys.} {\bf 344}  (1999) 573.

\bibitem{Just2016}
O.~{Just}, M.~{Obergaulinger}, H.-T. {Janka}, A.~{Bauswein} and N.~{Schwarz},
  {\em Astrophys. J. Letters} {\bf 816} (January 2016)   L30.

\bibitem{Perego2017}
A.~{Perego}, H.~{Yasin} and A.~{Arcones}, {\em J. Phys. G Nucl. Phys.} {\bf 44}
  (August 2017)   084007.

\bibitem{Hotokezaka:2013:24001}
K.~{Hotokezaka}, K.~{Kiuchi}, K.~{Kyutoku}, H.~{Okawa}, Y.-i. {Sekiguchi},
  M.~{Shibata} and K.~{Taniguchi}, {\em Phys. Rev. D} {\bf 87} (January 2013)
  024001.

\bibitem{Ciolfi2017}
R.~{Ciolfi}, W.~{Kastaun}, B.~{Giacomazzo}, A.~{Endrizzi}, D.~M. {Siegel} and
  R.~{Perna}, {\em Phys. Rev. D} {\bf 95} (March 2017)   063016.

\bibitem{Dessart:2009:1681}
L.~{Dessart}, C.~D. {Ott}, A.~{Burrows}, S.~{Rosswog} and E.~{Livne}, {\em
  Astrophys. J.} {\bf 690} (January 2009) 1681.

\bibitem{Siegel:2014:6}
D.~M. {Siegel}, R.~{Ciolfi} and L.~{Rezzolla}, {\em Astrophys. J. Letter} {\bf
  785} (April 2014)  ~L6.

\bibitem{Siegel2017}
D.~M. {Siegel} and B.~D. {Metzger}, {\em Phys. Rev. Letters} {\bf 119}
  (December 2017)   231102.

\bibitem{Meier2003}
D.~L. {Meier}, {\em New Astron. Rev.} {\bf 47} (October 2003) 667.

\bibitem{Abramowicz2013}
M.~A. {Abramowicz} and P.~C. {Fragile}, {\em Liv. Rev. Rel.} {\bf 16} (January
  2013)  ~1.

\bibitem{Thorne1986}
K.~S. {Thorne}, R.~H. {Price} and D.~A. {MacDonald}, {\em {Black holes: The
  membrane paradigm}} 1986.

\bibitem{Tchekhovskoy2012}
A.~{Tchekhovskoy} and J.~C. {McKinney}, {\em Mon. Not. R. Astron. Soc. Letters}
  {\bf 423} (June 2012) L55.

\bibitem{Shibata2011}
M.~{Shibata} and K.~{Taniguchi}, {\em Liv. Rev. Rel.} {\bf 14} (August 2011)
  ~6.

\bibitem{East2015}
W.~E. {East}, V.~{Paschalidis} and F.~{Pretorius}, {\em Astrophys. J. Letters}
  {\bf 807} (July 2015)  ~L3.

\bibitem{Chawla2010}
S.~{Chawla}, M.~{Anderson}, M.~{Besselman}, L.~{Lehner}, S.~L. {Liebling},
  P.~M. {Motl} and D.~{Neilsen}, {\em Phys. Rev. Letters} {\bf 105} (September
  2010)   111101.

\bibitem{Etienne2012a}
Z.~B. {Etienne}, Y.~T. {Liu}, V.~{Paschalidis} and S.~L. {Shapiro}, {\em Phys.
  Rev. D} {\bf 85} (March 2012)   064029.

\bibitem{Etienne2012b}
Z.~B. {Etienne}, V.~{Paschalidis} and S.~L. {Shapiro}, {\em Phys. Rev. D} {\bf
  86} (October 2012)   084026.

\bibitem{KiuchiSek2015}
K.~{Kiuchi}, Y.~{Sekiguchi}, K.~{Kyutoku}, M.~{Shibata}, K.~{Taniguchi} and
  T.~{Wada}, {\em Phys. Rev. D} {\bf 92} (September 2015)   064034.

\bibitem{Paschalidis2015}
V.~{Paschalidis}, M.~{Ruiz} and S.~L. {Shapiro}, {\em Astrophys. J. Letter}
  {\bf 806} (June 2015)   L14.

\bibitem{Balbus:1991}
S.~A. {Balbus} and J.~F. {Hawley}, {\em Astrophys. J.} {\bf 376} (July 1991)
  214.

\bibitem{Faber2012}
J.~A. Faber and F.~A. Rasio, {\em Liv. Rev. Rel.} {\bf 15}  (2012).

\bibitem{Rezzolla2011}
L.~{Rezzolla}, B.~{Giacomazzo}, L.~{Baiotti}, J.~{Granot}, C.~{Kouveliotou} and
  M.~A. {Aloy}, {\em Astrophys. J. Letters} {\bf 732} (May 2011)  ~L6.

\bibitem{Kiuchi:2014:41502}
K.~{Kiuchi}, K.~{Kyutoku}, Y.~{Sekiguchi}, M.~{Shibata} and T.~{Wada}, {\em
  Phys. Rev. D} {\bf 90} (August 2014)   041502.

\bibitem{Dionysopoulou:2015:92}
K.~{Dionysopoulou}, D.~{Alic} and L.~{Rezzolla}, {\em Phys. Rev. D} {\bf 92}
  (October 2015)   084064.

\bibitem{Kawamura:2016:064012}
T.~Kawamura, B.~Giacomazzo, W.~Kastaun, R.~Ciolfi, A.~Endrizzi, L.~Baiotti and
  R.~Perna, {\em Phys. Rev. D} {\bf 94} (Sep 2016)   064012.

\bibitem{Ruiz2016}
M.~{Ruiz}, R.~N. {Lang}, V.~{Paschalidis} and S.~L. {Shapiro}, {\em Astrophys.
  J. Letters} {\bf 824} (June 2016)  ~L6.

\bibitem{Rasio1999}
F.~A. {Rasio} and S.~L. {Shapiro}, {\em Class. Quantum Grav.} {\bf 16} (June
  1999) R1.

\bibitem{Price2006}
D.~J. {Price} and S.~{Rosswog}, {\em Science} {\bf 312} (May 2006) 719.

\bibitem{Siegel:2013:121302}
D.~M. {Siegel}, R.~{Ciolfi}, A.~I. {Harte} and L.~{Rezzolla}, {\em Phys. Rev. D
  R} {\bf 87} (June 2013)   121302.

\bibitem{Palenzuela2015}
C.~{Palenzuela}, S.~L. {Liebling}, D.~{Neilsen}, L.~{Lehner}, O.~L.
  {Caballero}, E.~{O'Connor} and M.~{Anderson}, {\em Phys. Rev. D} {\bf 92}
  (August 2015)   044045.

\bibitem{Zrake2013}
J.~{Zrake} and A.~I. {MacFadyen}, {\em Astrophys. J.} {\bf 769} (June 2013)
  L29.

\bibitem{GiacomazzoSub2015}
B.~Giacomazzo, J.~Zrake, P.~Duffell, A.~I. MacFadyen and R.~Perna, {\em
  Astrophys. J.} {\bf 809}  (2015)  ~39.

\bibitem{Radice2017}
D.~{Radice}, {\em Astrophys. J. Letters} {\bf 838} (March 2017)  ~L2.

\bibitem{Gehrels2004-Swift}
N.~{Gehrels}, G.~{Chincarini}, P.~{Giommi}, K.~O. {Mason}, J.~A. {Nousek},
  A.~A. {Wells}, N.~E. {White}, S.~D. {Barthelmy}, D.~N. {Burrows}, L.~R.
  {Cominsky}, et al., {\em Astrophys. J.} {\bf 611} (August 2004) 1005.

\bibitem{Norris2006}
J.~P. {Norris} and J.~T. {Bonnell}, {\em Astrophys. J.} {\bf 643} (May 2006)
  266.

\bibitem{Gompertz2013}
B.~P. {Gompertz}, P.~T. {O'Brien}, G.~A. {Wynn} and A.~{Rowlinson}, {\em Mon.
  Not. R. Astron. Soc.} {\bf 431} (May 2013) 1745.

\bibitem{Barthelmy2005b}
S.~D. {Barthelmy}, J.~K. {Cannizzo}, N.~{Gehrels}, G.~{Cusumano}, V.~{Mangano},
  P.~T. {O'Brien}, S.~{Vaughan}, B.~{Zhang}, D.~N. {Burrows}, S.~{Campana}, et al., 
  {\em Astrophysical J. Letters} {\bf 635} (December 2005) L133.

\bibitem{Campana2006}
S.~{Campana}, G.~{Tagliaferri}, D.~{Lazzati}, G.~{Chincarini}, S.~{Covino},
  K.~{Page}, P.~{Romano}, A.~{Moretti}, G.~{Cusumano}, V.~{Mangano}, et al., 
  {\em Astron. Astrophys.} {\bf 454} (July 2006) 113.

\bibitem{Rowlinson2010}
A.~{Rowlinson}, P.~T. {O'Brien}, N.~R. {Tanvir}, B.~{Zhang}, P.~A. {Evans},
  N.~{Lyons}, A.~J. {Levan}, R.~{Willingale}, K.~L. {Page}, O.~{Onal}, et al., 
  {\em Mon. Not. R. Astron. Soc.} {\bf 409} (December 2010) 531.

\bibitem{Rowlinson2013}
A.~{Rowlinson}, P.~T. {O'Brien}, B.~D. {Metzger}, N.~R. {Tanvir} and A.~J.
  {Levan}, {\em Mon. Not. R. Astron. Soc.} {\bf 430} (April 2013) 1061.

\bibitem{Lue2015}
H.-J. {L{\"u}}, B.~{Zhang}, W.-H. {Lei}, Y.~{Li} and P.~D. {Lasky}, {\em
  Astrophys. J.} {\bf 805} (June 2015)  ~89.

\bibitem{Rosswog2007}
S.~{Rosswog}, {\em Mon. Not. R. Astron. Soc.} {\bf 376} (March 2007) L48.

\bibitem{Zhang2001}
B.~{Zhang} and P.~{M{\'e}sz{\'a}ros}, {\em Astrophys. J. Letters} {\bf 552}
  (May 2001) L35.

\bibitem{Gao2006}
W.-H. {Gao} and Y.-Z. {Fan}, {\em Chinese J. Astron. Astrophys.} {\bf 6}
  (October 2006) 513.

\bibitem{Metzger2008}
B.~D. {Metzger}, E.~{Quataert} and T.~A. {Thompson}, {\em Mon. Not. R. Astron.
  Soc.} {\bf 385} (April 2008) 1455.

\bibitem{Rowlinson2014}
A.~{Rowlinson}, B.~P. {Gompertz}, M.~{Dainotti}, P.~T. {O'Brien}, R.~A.~M.~J.
  {Wijers} and A.~J. {van der Horst}, {\em Mon. Not. R. Astron. Soc.} {\bf 443}
  (September 2014) 1779.

\bibitem{Lasky2014}
P.~D. {Lasky}, B.~{Haskell}, V.~{Ravi}, E.~J. {Howell} and D.~M. {Coward}, {\em
  Phys. Rev. D} {\bf 89} (February 2014)   047302.

\bibitem{Yu2013}
Y.-W. {Yu}, B.~{Zhang} and H.~{Gao}, {\em Astrophys. J. Letters} {\bf 776}
  (October 2013)   L40.

\bibitem{Metzger2014b}
B.~D. {Metzger} and A.~L. {Piro}, {\em Mon. Not. R. Astron. Soc.} {\bf 439}
  (April 2014) 3916.

\bibitem{Siegel:2016a}
D.~M. {Siegel} and R.~{Ciolfi}, {\em Astrophys. J.} {\bf 819} (March 2016)
  14S.

\bibitem{Siegel:2016b}
D.~M. {Siegel} and R.~{Ciolfi}, {\em Astrophys. J.} {\bf 819} (March 2016)
  15S.

\bibitem{Ciolfi2016}
R.~{Ciolfi}, {\em Astrophys. J.} {\bf 829} (October 2016)  ~72.

\bibitem{THESEUS-WP}
L.~{Amati}, P.~{O'Brien}, D.~{Goetz}, E.~{Bozzo}, C.~{Tenzer}, F.~{Frontera},
  G.~{Ghirlanda}, C.~{Labanti}, J.~P. {Osborne}, G.~{Stratta}, et al., 
  {\em ArXiv e-prints}  (October 2017).

\bibitem{THESEUS-MM}
G.~{Stratta}, R.~{Ciolfi}, L.~{Amati}, G.~{Ghirlanda}, N.~{Tanvir}, E.~{Bozzo},
  D.~{Gotz}, P.~{O'Brien}, F.~{Frontera}, J.~P. {Osborne}, et al., 
  {\em ArXiv e-prints}  (December 2017).

\bibitem{Duez2006b}
M.~D. {Duez}, Y.~T. {Liu}, S.~L. {Shapiro}, M.~{Shibata} and B.~C. {Stephens},
  {\em Phys. Rev. D} {\bf 73} (May 2006)   104015.

\bibitem{Kiuchi:2012:86}
K.~{Kiuchi}, K.~{Kyutoku} and M.~{Shibata}, {\em Phys. Rev. D} {\bf 86}
  (September 2012)   064008.

\bibitem{GiacomazzoPerna}
B.~{Giacomazzo} and R.~{Perna}, {\em Astrophys. J. Letters} {\bf 771} (July
  2013)   L26.

\bibitem{Endrizzi:2016:164001}
A.~Endrizzi, R.~Ciolfi, B.~Giacomazzo, W.~Kastaun and T.~Kawamura, {\em Class.
  Quantum Grav.} {\bf 33}  (2016)   164001.

\bibitem{Ciolfi:2015:36}
R.~{Ciolfi} and D.~M. {Siegel}, {\em Astrophys. J. Letters} {\bf 798} (January
  2015)   L36.

\bibitem{Troja2010}
E.~{Troja}, S.~{Rosswog} and N.~{Gehrels}, {\em Astrophys. J.} {\bf 723}
  (November 2010) 1711.

\bibitem{Rezzolla2015}
L.~{Rezzolla} and P.~{Kumar}, {\em Astrophys. J.} {\bf 802} (April 2015)  ~95.

\bibitem{Margalit2015}
B.~{Margalit}, B.~D. {Metzger} and A.~M. {Beloborodov}, {\em Phys. Rev.
  Letters} {\bf 115} (October 2015)   171101.

\bibitem{Shibata2000}
M.~{Shibata}, T.~W. {Baumgarte} and S.~L. {Shapiro}, {\em Phys. Rev. D} {\bf
  61} (February 2000)   044012.

\bibitem{Baiotti:2005:24035}
L.~Baiotti, I.~Hawke, P.~J. Montero, F.~L{\"o}ffler, L.~Rezzolla,
  N.~Stergioulas, J.~A. Font and E.~Seidel, {\em Phys. Rev. D} {\bf 71}  (2005)
    024035.

\bibitem{Kastaun:2015:064027}
W.~Kastaun and F.~Galeazzi, {\em Phys. Rev. D} {\bf 91} (Mar 2015)   064027.

\bibitem{Kastaun:2016}
W.~{Kastaun}, R.~{Ciolfi} and B.~{Giacomazzo}, {\em Phys. Rev. D} {\bf 94}
  (August 2016)   044060.

\bibitem{Kastaun2017}
W.~{Kastaun}, R.~{Ciolfi}, A.~{Endrizzi} and B.~{Giacomazzo}, {\em Phys. Rev.
  D} {\bf 96} (August 2017)   043019.

\bibitem{Hanauske2017}
M.~{Hanauske}, K.~{Takami}, L.~{Bovard}, L.~{Rezzolla}, J.~A. {Font},
  F.~{Galeazzi} and H.~{St{\"o}cker}, {\em Phys. Rev. D} {\bf 96} (August 2017)
    043004.

\bibitem{Pringle1981}
J.~E. {Pringle}, {\em Ann. Rev. Astron. Astrophys.} {\bf 19}  (1981) 137.

\bibitem{Shapiro2000}
S.~L. {Shapiro}, {\em Astrophys. J.} {\bf 544} (November 2000) 397.

\bibitem{Hjorth2017}
J.~{Hjorth}, A.~J. {Levan}, N.~R. {Tanvir}, J.~D. {Lyman}, R.~{Wojtak}, S.~L.
  {Schr{\o}der}, I.~{Mandel}, C.~{Gall} and S.~H. {Bruun}, {\em Astrophys. J.
  Letters} {\bf 848} (October 2017)   L31.

\bibitem{Im2017}
M.~{Im}, Y.~{Yoon}, S.-K.~J. {Lee}, H.~M. {Lee}, J.~{Kim}, C.-U. {Lee}, S.-L.
  {Kim}, E.~{Troja}, C.~{Choi}, G.~{Lim}, et al., 
  {\em Astrophys. J. Letters} {\bf 849} (November 2017)   L16.

\bibitem{DAvanzo2018}
P.~{D'Avanzo}, S.~{Campana}, G.~{Ghisellini}, A.~{Melandri}, M.~G.
  {Bernardini}, S.~{Covino}, V.~{D'Elia}, L.~{Nava}, R.~{Salvaterra},
  G.~{Tagliaferri} and S.~D. {Vergani}, {\em ArXiv e-prints}  (January 2018).

\bibitem{Salafia2018}
O.~S. {Salafia}, G.~{Ghisellini}, G.~{Ghirlanda} and M.~{Colpi}, {\em ArXiv
  e-prints}  (November 2017).

\bibitem{LVC-Hubble}
B.~P. {Abbott}, R.~{Abbott}, T.~D. {Abbott}, F.~{Acernese}, K.~{Ackley},
  C.~{Adams}, T.~{Adams}, P.~{Addesso}, R.~X. {Adhikari}, V.~B. {Adya}, et al., 
  {\em Nature} {\bf 551} (November 2017) 85.

\bibitem{Margalit2017}
B.~{Margalit} and B.~D. {Metzger}, {\em Astrophys. J. Letters} {\bf 850}
  (December 2017)   L19.

\bibitem{Shibata2017}
M.~{Shibata}, S.~{Fujibayashi}, K.~{Hotokezaka}, K.~{Kiuchi}, K.~{Kyutoku},
  Y.~{Sekiguchi} and M.~{Tanaka}, {\em Phys. Rev. D} {\bf 96} (December 2017)
  123012.

\bibitem{Radice2018}
D.~{Radice}, A.~{Perego}, F.~{Zappa} and S.~{Bernuzzi}, {\em Astrophys. J.
  Letters} {\bf 852} (January 2018)   L29.

\bibitem{Sugita2017}
S.~{Sugita}, N.~{Kawai}, M.~{Serino} and et~al., {\em GCN, 21555}   (2017).

\bibitem{Evans2017}
P.~A. {Evans}, S.~B. {Cenko}, J.~A. {Kennea}, S.~W.~K. {Emery}, N.~P.~M.
  {Kuin}, O.~{Korobkin}, R.~T. {Wollaeger}, C.~L. {Fryer}, K.~K. {Madsen},
  F.~A. {Harrison}, et al., {\em Science} {\bf 358} (December 2017) 1565.

\end{thebibliography}


\end{document}